\documentclass[lettersize,journal]{IEEEtran}
\usepackage{amsmath,amsfonts}

\usepackage{algorithmic}
\usepackage{algorithm}

\usepackage{array}
\usepackage[caption=false,font=footnotesize,labelfont=rm,textfont=rm]{subfig}
\usepackage{textcomp}
\usepackage{tabularx}
\usepackage{stfloats}
\usepackage{setspace}
\usepackage{url}
\usepackage{verbatim}
\usepackage{graphicx}
\usepackage{hyperref}
\usepackage{cite}
\usepackage{makecell}
\usepackage{xcolor}      
\usepackage{colortbl}    
\usepackage{booktabs}    
\usepackage{microtype}
\usepackage{nccmath}
\usepackage{mathtools}
\begin{document}

	\title{Data-Model Co-Driven Continuous Channel Map Construction: A Perceptive Foundation for Embodied Intelligent Agents in 6G Networks}

	\author{Tianrun Qi,~\IEEEmembership{Student Member,~IEEE}, Cheng-Xiang Wang,~\IEEEmembership{Fellow,~IEEE}, Chen Huang,~\IEEEmembership{Member,~IEEE}, \\Junling Li,~\IEEEmembership{Member,~IEEE}, and John S Thompson,~\IEEEmembership{Fellow,~IEEE}.
		\thanks{This work was supported by the National Natural Science Foundation of China (NSFC) under Grants 62394290, 62394291,and 62394294, the Research Fund of National Mobile Communications Research Laboratory, Southeast University, under Grant 2026A05, the Natural Science Foundation of Jiangsu Province under Grant BK20243059, the China Scholarship Council (CSC) under Grant 202506090282, and the UK Engineering and Physical Sciences Research Council (EPSRC) under Grants EP/X04047X/2 and EP/Y037243/1 for the TITAN Telecoms Hub. (\emph{Corresponding Authors: Cheng-Xiang Wang and Chen Huang})}
		\IEEEcompsocitemizethanks{
			\IEEEcompsocthanksitem T. Qi is with the National Mobile Communications Research Laboratory, School of Information Science and Engineering, Southeast University, Nanjing 210096, China (email: \{tr\_qi\}@seu.edu.cn).
			\IEEEcompsocthanksitem C.-X. Wang and J. Li are with the National Mobile Communications Research Laboratory, School of Information Science and Engineering, Southeast University, Nanjing 211189, China, and also with the Pervasive Communication Research Center, Purple Mountain Laboratories, Nanjing 211111, China (e-mail: \{chxwang, junlingli\}@seu.edu.cn).
			\IEEEcompsocthanksitem C. Huang is with the Pervasive Communication Research Center, Purple Mountain Laboratories, Nanjing 211111, China, and also with the National Mobile Communications Research Laboratory, School of Information Science and Engineering, Southeast University, Nanjing 210096, China (e-mail: huangchen@pmlabs.com.cn). 
			\IEEEcompsocthanksitem John S Thompson is with the Institute for Imaging, Data and Communications, School of Engineering, University of Edinburgh, EH9 3BF Edinburgh, U.K. (e-mail: john.thompson@ed.ac.uk). 
			
	}}

	\maketitle

	\begin{abstract}

Future 6G networks will host massive numbers of embodied intelligent agents, which require real-time channel awareness over continuous-space for autonomous decision-making. By pre-obtaining location-specific channel state information (CSI), channel map can be served as a foundational world model for embodied intelligence to achieve wireless channel perception. 
However, acquiring CSI via measurements is costly, so in practice only sparse observations are available, leaving agents “blind” to channel conditions at unvisited locations. Meanwhile, purely model-driven channel maps can provide dense CSI but often yields unsatisfactory accuracy and robustness, while purely data-driven interpolation from sparse measurements 
is computationally prohibitive for real-time updates.
To address these challenges, this paper proposes a data–model co-driven (DMcD) framework that performs a two-stage interpolation toward a space-time continuous channel map, 
First, a hybrid ray tracing and geometry-based channel model (H-RT/GBSM) is developed to capture dynamic scatterers, providing dense, time-variant channel properties that match measurement statistics as a physically consistent prior. Then, an inductive edge-conditioned graph neural network (InductE-GNN) fuses the prior with sparse measurements to perform real-time spatial interpolation, enabling rapid online adaptation without retraining, ensuring the synchronization with the dynamic physical reality.
Evaluations with measured datasets show that the proposed DMcD framework significantly outperforms data-only and model-only baselines, providing accurate and queryable channel information for embodied intelligent agents.

	\end{abstract}
\vspace{-0.5em}
	\begin{IEEEkeywords}
		6G, embodied intelligent agent, data–model co-driven framework, space-time continuous channel map construction, online spatial interpolation.
	\end{IEEEkeywords}
\vspace{-1em}
	\section{Introduction}

	In recent years, the sixth-generation (6G) communication system has been widely discussed as an artificial intelligence(AI)-native communication network \cite{backgroud_CXWang,backgroud_ZhangRuichen}. Instead of only carrying traffic, the network is expected to host distributed intelligence and to support autonomous decision making in real time \cite{Letaief2022EdgeAI6G,Li20_TCCN_DRL_EdgeVehicular}. On top of this AI-native infrastructure, networked agents will play an important role. In this paper, the agents can be autonomous unmanned aerial vehicles (UAVs), connected vehicles, service robots, and other smart devices with computation and communication modules. These embodied intelligent agents move in the physical world, sense their environment, and interact with the wireless network. Obviously, they need to make many networking decisions by themselves, such as selecting access points and beams, planning trajectories, and processing integrated sensing and communication (ISAC) tasks \cite{Liu2022ISAC,Heng21_TCCN_MLBeamAlign}. To provide such capabilities, the agent should not only react to the collected static channel state, but also require the real-time channel information in continuous space. This calls for some form of world model \cite{backgroud_ZhangRuichen_2,WorldModels_3} of the wireless environment that can be queried and updated at run time.
	
	However, most current wireless communication systems still rely on local instantaneous channel state information (CSI), which is obtained by pilot-based channel estimation or beam training at a single time and location, then used for link adaptation and beam management \cite{CKM_1}. Such a system works well in static or slowly-varying scenarios but is not sufficient for agent decisions, since it requires dense measurements or estimation to track the time-variations of continuous-space channels, further leading to excessive pilot overhead and latency \cite{COMST25,EIT2}. Therefore, agents often have to make decisions based on discrete and out-of-date CSI, without a consistent model linking different locations and time instants together\cite{Li20_TCCN_DRL_EdgeVehicular}.

	Channel maps have been proposed as a way to overcome this limitation. Channel maps describe location-specific channel characteristics over a geographic area\cite{CKM_1}. Unlike conventional radio maps that only provide large scale CSI like path loss, received power \cite{radiodiff,Xu21_TCCN_BayesianREM}, channel maps can give both large-scale and small-scale channel information.
	Once constructed, it allows the network to infer channel quality at unmeasured locations and to support coverage planning, localization, and other environment-aware functions \cite{QTR_TVT,CKM_1}. Existing work on channel maps have shown clear benefits for resource management and environment-aware communications \cite{Chen2025JSTSP,Zeng21,Zhang2025UAV}.
	Yet, most existing channel maps are either time-invariant or constructed in discrete-space. They rarely provide a truly spatio-temporal representation that follows the motion of scatterers. This gap becomes critical when serving embodied intelligent agents that move quickly and require continuous-space channel information along their trajectories.
	
	Channel maps have emerged as a critical enabler for environment-aware 6G networks, assisting embodied intelligent agents in making informed decisions for diverse tasks. For instance,  a channel map-based angle domain multiple access scheme was introduced in~\cite{Chen2025JSTSP}, utilizing the constructed angle map to acquire angle-of-departure information for efficient user grouping and beamforming. Furthermore, channel maps was strategically applied to UAV networks in~\cite{Zhang2025UAV}, designing a trajectory planning scheme that minimizes outage probability and propulsion energy consumption. Beyond trajectory design, channel maps were also utilized to compensate for positioning errors in dense urban scenarios~\cite{Zhang2024Pos}, where map-assisted schemes improved the reliability of location-based services. These applications underscore the potential of channel maps as a perceptive foundation for embodied intelligent agents.
	
	Current advanced channel map construction methods can be broadly classified into two categories: Model-driven methods and data-driven methods. Model-driven methods leverage stochastic channel models to obtain continuous-space channel information based on physical propagation environments~\cite{Wang6GPCM3,Wang6GPCM2,Zhu2022Map,Wang2024SBL,Sun2025Gain,YangYueEIT,QTR_TVT}. Among these, the geometry-based stochastic model (GBSM) is widely adopted to capture the statistical behavior of wireless channels across various scenarios~\cite{Wang6GPCM3,Wang6GPCM2}.
	To address the specific challenges of vehicle-to-vehicle communications, a map-based channel modeling method was proposed for UAV-to-vehicle links in \cite{Zhu2022Map}, incorporating 3D environment reconstruction to generate realistic channel realizations considering dense multipath components (DMCs).
	A hierarchical construction framework based on sparse Bayesian learning was introduced in \cite{Wang2024SBL}, which mathematically decomposed the channel map into path loss, shadowing, and multipath components (MPCs) to enhance modeling precision.
	Recently, a channel gain map estimation method was developed based on a scatterer model in \cite{Sun2025Gain}, where scatterer locations and reflection coefficients were identified to assist Kriging interpolation in constructing the gain map.
	In addition to stochastic approaches, electromagnetic (EM) field-based models were extensively studied \cite{EIT2,YangYueEIT}. In \cite{YangYueEIT}, a 3D continuous-space EM channel model incorporating scatterers and spherical wavefronts was proposed and accurately calculated scattering fields using the method of moments.

	Data-driven methods construct channel maps by learning the underlying mapping relationships from collected discrete channel data. Early approaches typically utilized mathematical interpolation techniques\cite{Kriging,Xu21_TCCN_BayesianREM}, such as Kriging~\cite{Kriging}, to estimate channels at unmeasured locations. With the advent of AI, deep learning offered a new paradigm for extracting non-linear spatial features from measurement data \cite{Wu2024CKMImageNet,Wang2025SR,channelGAN,Hu2025,GNN2,GNN3,QTR_TCOM,HuangTAP1,HuangTAP2}.  Neural networks like the gated recurrent unit (GRU) and long short-term memory networks were applied to capture spatial correlations [5]. In~\cite{HuangTAP2}, a hybrid Conv-GRU algorithm that enables frequency-domain channel reconstruction by learning from angular power spectra was proposed. In~\cite{Wu2024CKMImageNet}, a comprehensive CKM dataset was constructed to facilitate the training of data-driven models for tasks like channel estimation and beam prediction. To further improve the resolution of constructed maps, a deep learning-based framework utilizing image super-resolution technology was proposed in~\cite{Wang2025SR}. 
	Generative AI has also been explored to address data sparsity, including generative adversarial networks (GANs) and diffusion algorithms. a Conditional GAN (CGAN)-based framework was proposed in~\cite{channelGAN,Hu2025,radiodiff}, where the generator adopted a U-Net structure to jointly construct THz radio maps and detect obstacles. In~\cite{radiodiff}, the sampling-free radio map construction problem is modeled as a conditional generative task and a denoising diffusion method (RadioDiff) is proposed.
	Nevertheless, most existing AI-enabled channel map construction methods focus on interpolating \emph{a single channel characteristic} (e.g., pathloss or angle-related attributes) rather than recovering the \emph{complete high-dimensional channel matrix}.
	As a result, the ``fast prediction'' claimed in related works typically provides only partial large-scale channel parameters, which remains insufficient to effectively support embodied intelligent agents in decision-making that requires fine-grained, full-CSI awareness.
	Recently, Graph Neural Networks (GNNs) and Transformer architectures have emerged as powerful tools for characterizing the topological structures of wireless networks \cite{GNN2,GNN3}. In~\cite{GNN2}, channel map prediction was formulated as a graph-based learning task and proposed a graph convolutional network (GCN) with filters to estimate power spectral density (PSD) at unmeasured locations using discrete measurement nodes as graph inputs. In \cite{GNN3}, a two-stage method combining environment graph construction with GNN models was proposed.

\begin{figure}[t]
	\centering
	\includegraphics[width=9.3cm]{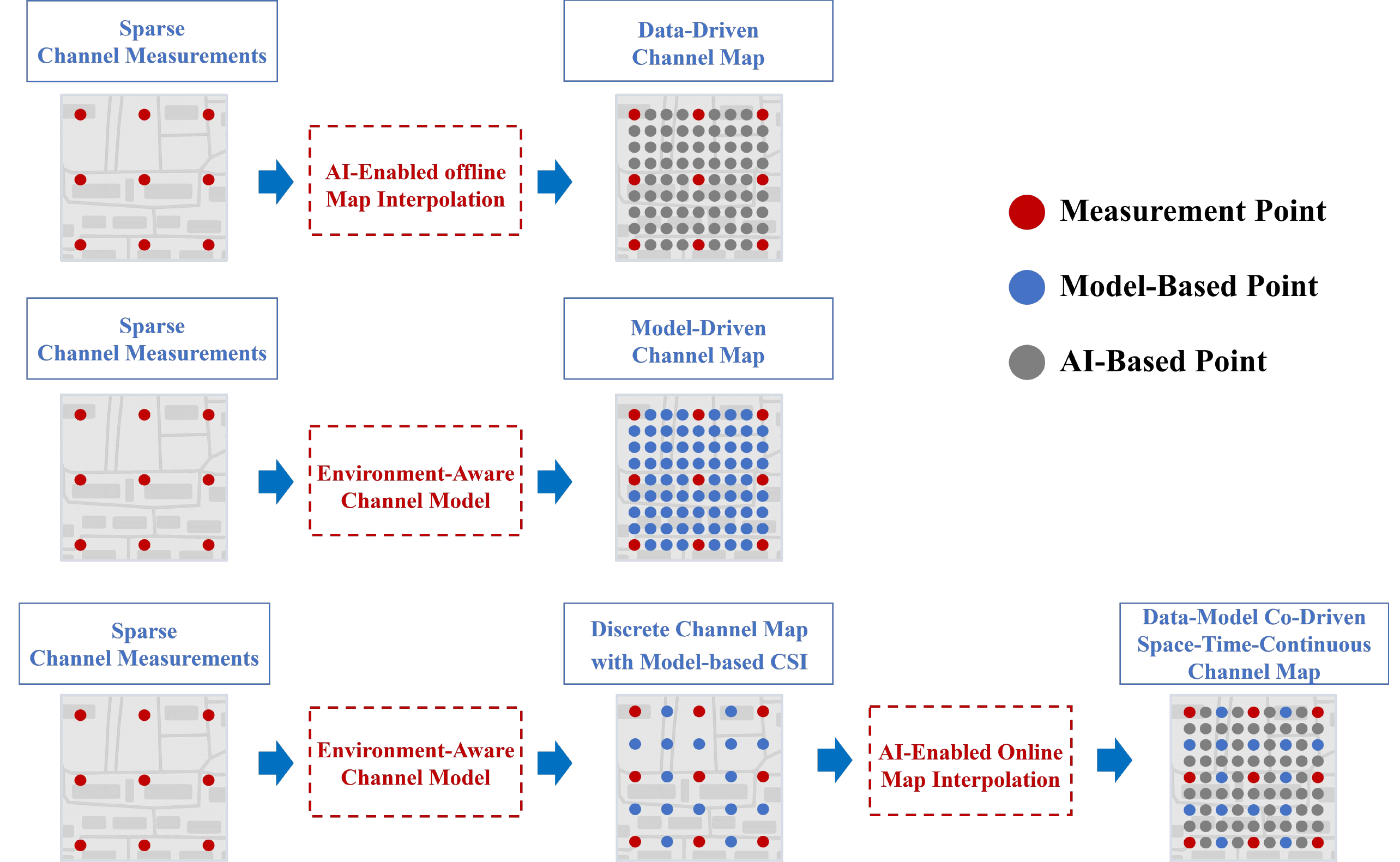}
	\caption{Comparison of channel map construction paradigms.}
	\label{fig:dmcd_concept}
\end{figure}
	Summarizing the existing approaches mentioned above, data-driven channel maps usually achieve high accuracy in the continuous space, but they cannot follow fast channel variations because the AI algorithms need to be retrained once the environment significantly changes. In contrast, model-driven channel maps can adapt to real-time variations of environment, but their stochastic nature limits the location-specific accuracy. This gap motivates a joint approach for space-time continuous channel maps rather than relying on either data-driven or model-driven methods alone.

	To fill the gap of the joint approach, this paper proposes a Data-Model Co-Driven (DMcD) method for space-time continuous channel map construction. Fig.~\ref{fig:dmcd_concept} illustrates the core motivation of this work.
	Our key viewpoint is to generate the dynamic channel information by a channel model, enabling more accurate learning and AI-enabled interpolation with a larger and more reasonable training set. To the best of authors’ knowledge, this is the first time that a data-model co-driven framework for channel map construction has been proposed.

		\begin{figure*}[!b]
	\centering
	\includegraphics[width=17.5cm]{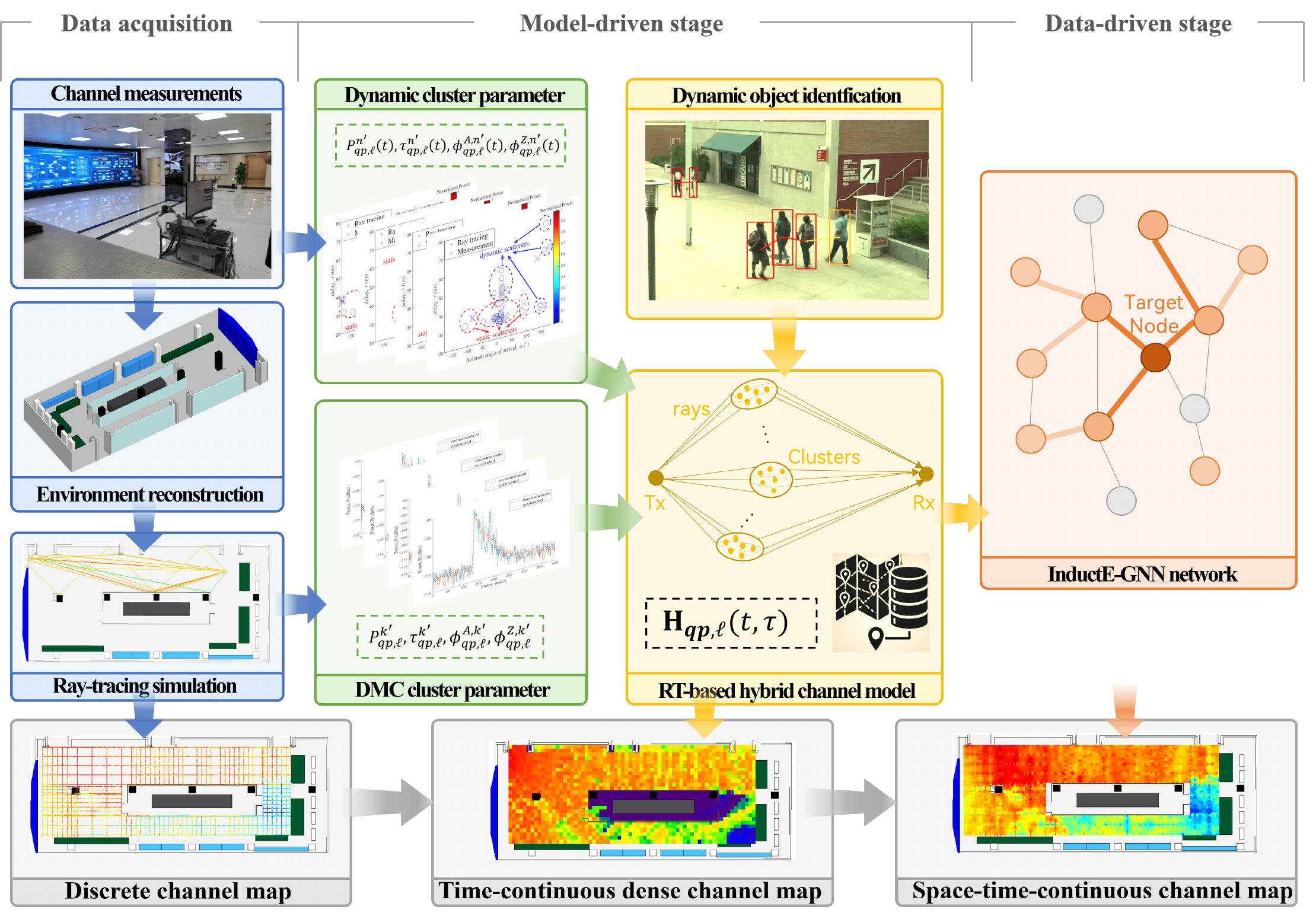}
	\caption{Two-stage data–model co-driven framework for space-time continuous channel map construction.}
	\label{fig:Framework}
\end{figure*}

	\begin{itemize}
		\item 
		The channel model and AI algorithm are integrated into a unified DMcD framework for space-time continuous channel map construction.
		The channel model provides dynamic channel information as priors, while the AI algorithm online interpolates the channel map from the priors and measurements.
		
		\item 
		A hybrid ray tracing and geometry-based channel model (H-RT/GBSM) is proposed to capture the impact of static scatterers, dynamic scatterers and DMCs. The hybrid channel model (HCM) describes the time evolution of scatterers and can generate real-time CSI that matches measurements in dynamic environments.
		
		\item 
		To achieve real-time construction, a novel inductive edge-conditioned graph neural network (InductE-GNN) is designed to interpolate the channel map in space. Unlike typical offline-trained AI algorithms whose performance degrades under distribution shift unless fine-tuned, the proposed inductE-GNN refreshes node representations at inference time, enabling re-training-free adaptation in dynamic channels.
		
		\item 
		The framework is evaluated on measured results against other AI algorithm baselines, and is shown to achieve higher accuracy and space-time continuous channel maps for embodied intelligent agents in 6G networks.
		
	\end{itemize}

	The rest of this paper is organized as follows. The two-stage DMcD framwork are presented in Section~II. Section~III gives a detailed description of data-acquisition stage. In Section~IV, the model-driven stage with the HCM are described. The data-driven stage and the InductE-GNN are given in Section~V. Section~VI gives a entire performance evaluation of HCM, InductE-GNN, and DMcD framework. Finally, conclusions are drawn in Section VII.

	\section{Data-Model Co-Driven Channel Map Construction Framework}

	In this section, the overall data-model co-driven framework for space-time continuous channel map construction is introduced. The whole procedure is divided into three parts, as illustrated in Fig.~\ref{fig:Framework}, including data acquisition, a model-driven stage, and a data-driven stage. The Fig.~\ref{fig:Framework} also shows the main input and output of each stage.

	1) Data acquisition and environment reconstruction:
	In the first stage, indoor and outdoor channel measurements are collected in the target environment by moving the receiver over a dense grid. The physical layout of the environment, including walls, furniture, and large objects, is then reconstructed from floor plans and on-site photos. A RT simulator is calibrated with this geometry and with the measured channel statistics. The RT simulation produces a discrete channel map on the measurement grid, which serves as the basis for the following model-driven stage.
	
	2) Model-driven stage with hybrid channel model:
	In the second stage, a RT-based HCM is built on top of the reconstructed environment. The HCM uses calibrated cluster parameters and explicitly considers static scatterers, dynamic scatterers, and DMCs. Dynamic objects such as moving people or robots are identified and mapped to time-varying clusters in the HCM. In this way, the model describes the time evolution of clusters and scatterers and generates time-continuous CSI over the area of interest. The output of this stage is a dense, time-continuous channel map that already reflects the physical dynamics of the physical scenario.
	
	3) Data-driven stage with InductE-GNN:
	In the third stage, the channel map is further refined in the spatial domain by a data-driven model. A graph is constructed on the channel map grid, where each node represents a location and carries channel features. The proposed InductE-GNN takes the model-driven map and a small set of new measurements as input, and performs online interpolation and correction on the graph. Thanks to its inductive design, the InductE-GNN can handle new nodes and updated measurements without retraining the whole algorithm. The final output is a space--time-continuous channel map that combines the physical consistency of the HCM with the accuracy of data-driven learning.
	
	In the next section, we discuss the datasets that this three stage DMCD approach is applied to.

	\section{Data Acquisition Stage of the Channel Map Construction}

	\subsection{Channel Measurements}
	To validate the proposed DMcD framework and calibrate the hybrid channel model, two channel measurement campaigns are conducted. One is a static indoor campaign at 28 GHz with rich DMCs, and the other is a dynamic outdoor campaign at 5.5 GHz with rich moving scatterers. Using both a mmWave band and a sub-6 GHz band helps to demonstrate the general applicability of the framework to different carrier frequencies and propagation conditions. In addition, the 28 GHz indoor scenario provides strong diffuse scattering, which is well suited for studying DMCs in this paper, while the 5.5 GHz outdoor scenario highlights the impact of dynamic objects on large-scale and small-scale channel properties. The detailed antenna array structures, calibration procedures, and full system parameters can be found in \cite{QTR_TVT,QTR_TCOM}.

	\subsubsection{Indoor static measurements with rich DMCs (28 GHz)}
	
	The first campaign is an ultra-dense continuous-space channel measurement in a typical indoor office environment at 28 GHz. The scenario is a rectangular hall of about 40 m × 20 m × 3.5 m, with glass windows, metal walls, and a metal ceiling. Inside the hall, a 15 m × 10 m inner room and a metal pillar create several non-line-of-sight (NLoS) regions and strong scattering clusters. A moving-sounder platform with synchronized transmitter (Tx) and receiver (Rx) is used. The Tx employs a single omnidirectional antenna at a height of about 3.0–3.3 m. The Rx is a 4×4 dual-polarized uniform planar array with 32 elements. During measurements, the Rx array is placed on a trolley and moved along 68 routes that cover both line-of-sight (LoS) and NLoS regions. In total, 11,974 spatial locations are recorded, forming an ultra-dense grid that approximates a continuous-space channel.
	These indoor data provide high-resolution delay PSD measurements with pronounced diffuse components. They are mainly used to identify and calibrate the DMC component of the HCM, and to test the proposed channel map construction under rich DMC conditions.

	\subsubsection{Outdoor dynamic measurements with rich moving scatterers (5.5 GHz)}
	
	The second campaign is an outdoor urban measurement at sub-6 GHz. The sounder operates at 5.5 GHz with 320 MHz bandwidth. The Tx uses a single omnidirectional antenna mounted on the eighth floor of an office building at a height of about 30 m. The Rx employs an 8×8 dual-polarized cylindrical array with 64 elements, mounted on a trolley that moves at street level. Multiple LoS and NLoS routes are designed between the buildings. At each route, the Rx trolley moves along the planned path while the sounder records multiple snapshots at each location to capture small-scale fading and Doppler effects caused by moving vehicles and pedestrians.
	This outdoor dataset is mainly used to characterize dynamic scatterers and to calibrate the time-varying part of the HCM. Together with the indoor campaign, it enables the proposed DMcD framework to be evaluated under both rich-DMC static conditions and rich-scatterer dynamic conditions, across mmWave and sub-6 GHz bands.
	
	The processing of raw sounding data, including back-to-back calibration, extraction of channel impulse responses, and high-resolution estimation of MPCs using the space-alternating generalized expectation-maximization (SAGE) algorithm, follows the procedures described in~\cite{III-1}, which is modified from~\cite{III-3} for mm-wave channels.
	
	\begin{figure}[!b]
		\centering
		\centering
		\subfloat[]
		{
			\hspace{-4mm}
			\includegraphics[width=4.5cm]{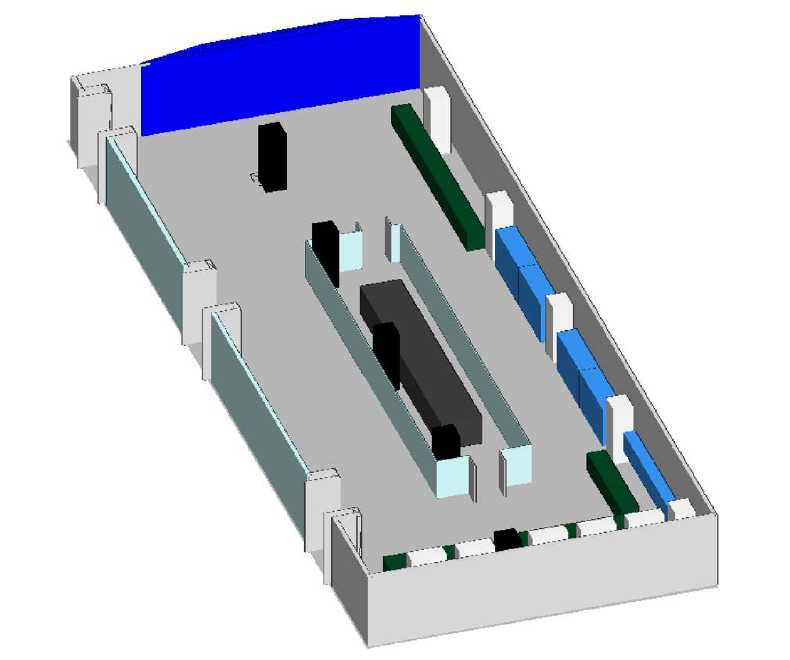}
			\label{fig:RT_indoor}
		}
		\subfloat[]
		{
			\hspace{-4mm}
			\includegraphics[width=4.5cm]{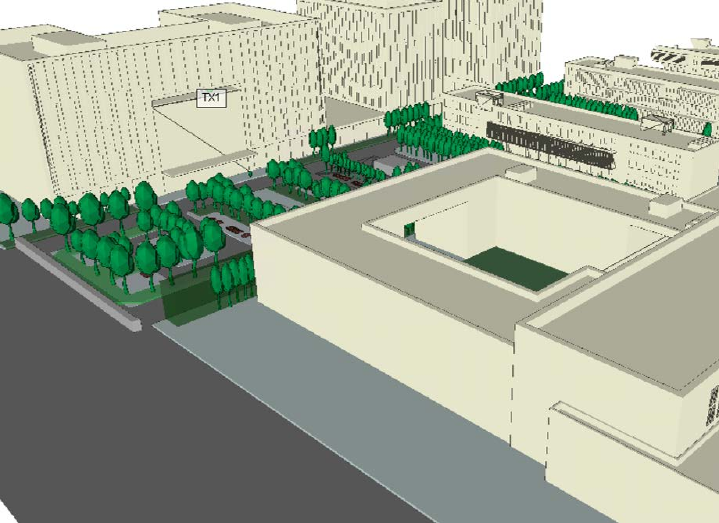}
			\label{fig:RT_outdoor}
		}
		
		\caption{Reconstruction of the (a) indoor and (b) outdoor measurement environments with RT simulations.}
		\label{fig:RT_Simulation}
	\end{figure}

	\subsection{Environment Reconstruction and Ray Tracing Simulation}

	RT is employed to reconstruct the static part of the propagation environment for both the indoor and outdoor scenarios. As shown in Fig.~\ref{fig:RT_Simulation}, the measurement environments are emulated in the commercial RT software Wireless InSite. For the indoor campaign, simulation points are placed on a dense grid with 1-meter spacing. Only static interaction objects, such as walls, pillars, and furniture, are included at this stage. The simulation frequency, bandwidth, and antenna configurations are kept consistent with the measurements. For the outdoor campaign, the China Network Valley scenario is also modeled in Wireless InSite following the layout of the real environment. The same carrier frequency of 5.5~GHz, bandwidth, and Tx/Rx antenna settings as in the measurements are adopted. LoS and reflected paths from buildings and trees are considered, while moving vehicles and pedestrians are not included in the RT scene and will be represented later by dynamic clusters in the HCM.
	
	The RT simulations provide delay, angle, and power parameters of specular multipath components (SMCs) on the simulation locations. These ray parameters are then used as inputs to construct the static component of the proposed HCM in Section~IV, and to determine the cluster centroids for DMCs.

		\section{ Model-driven stage of the Channel Map Construction}
	
	\begin{figure*}[t]
		\centering
		\includegraphics[width=16.5cm]{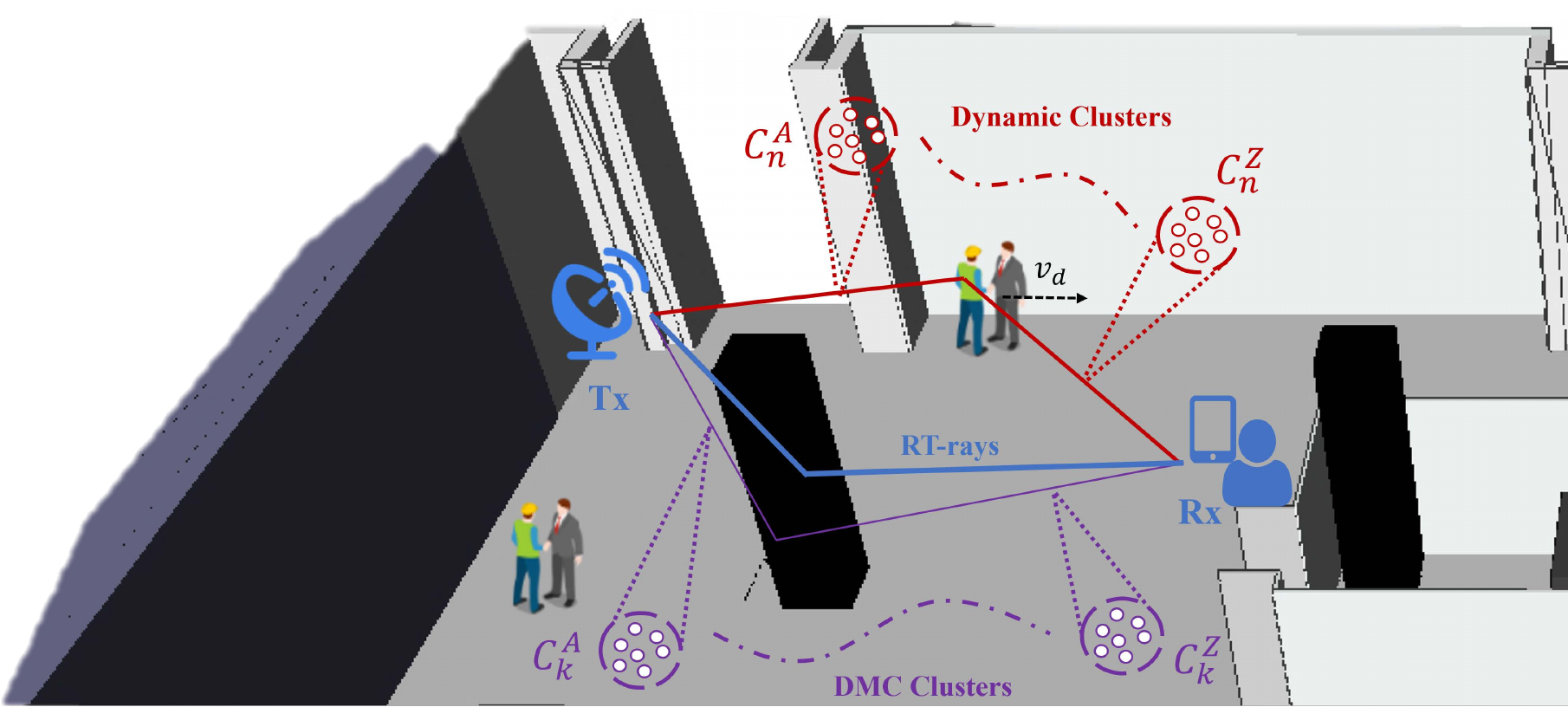}
		\caption{Geometry map of the proposed RT-based hybrid channel model.}
		\label{fig:GeometryMap_HCM}
	\end{figure*}

	The geometry of the proposed RT-based hybrid channel model is illustrated in Fig.~\ref{fig:GeometryMap_HCM}. The measurement scenario is reconstructed in 3D, where the Tx and Rx are placed according to the real deployment. The channel impulse response (CIR) consists with three components and can be calculated as
	\begin{equation}
		\begin{aligned}
		h_{qp, \ell}(t, \tau) &= \sqrt{\frac{1}{K_{\rm{D}}^{-1} + K_{\rm{DMC}}^{-1} + 1}} h_{qp, \ell}^{\rm{RT}}(\tau) \\
		&+ \sqrt{\frac{K_{\rm{D}}^{-1}}{K_{\rm{D}}^{-1} + K_{\rm{DMC}}^{-1} + 1}} h_{qp, \ell}^{\rm{D}}(t, \tau) \\
		&+ \sqrt{\frac{K_{\rm{DMC}}^{-1}}{K_{\rm{D}}^{-1} + K_{\rm{DMC}}^{-1} + 1}} h_{qp, \ell}^{\rm{DMC}}(\tau) 
		\end{aligned}
	\end{equation}
	where $K_{\rm{D}}$ and $K_{\rm{DMC}}$ are the power ratio of the RT paths to the dynamic multipaths and DMC, respectively. Additionally, $h_{qp, \ell}^{\rm{RT}}(\tau)$, $h_{qp, \ell}^{\rm{D}}(t, \tau)$, and $h_{qp, \ell}^{\rm{DMC}}(\tau)$ are the CIRs of specular static component, dynamic component, and DMC component.

        \subsection{Static and Dynamic Components of Hybrid Channel Model}
	Static SMCs obtained from RT are drawn as blue rays between the Tx and Rx. Dynamic clusters, shown in red, are associated with moving interaction objects such as pedestrians and vehicles. DMC clusters, shown in purple, are located near walls and other obstacles and will be discussed in the next subsection. The numbers of antennas at the Tx and Rx sides are denoted as $M_T$ and $M_R$, respectively. $A_q^T$ and $A_p^R$ denote the $q$-th ($q=1,\ldots,M_T$) Tx antenna element and the $p$-th ($p=1,\ldots,M_R$) Rx antenna element. Only the $k$-th ($k=1,\ldots,K_{qp,\ell}$) RT ray between $A_q^T$ and $A_p^R$ at location $\ell$ is illustrated in Fig.~\ref{fig:GeometryMap_HCM}.
	
	The static component of the HCM is represented by RT. For the link between $A_q^T$ and $A_p^R$ at location $\ell$, the RT-based CIR is written as \cite{QTR_TVT}
	\begin{equation}
		h_{qp,\ell}^{\rm{RT}}(\tau)=\sum_{k=1}^{K_{qp,\ell}}\sqrt{P_{qp,\ell}^k}\cdot e^{j2\pi f_c\tau_{qp,\ell}^k}\cdot\delta(\tau-\tau_{qp,\ell}^k) \label{h_rt}
	\end{equation}
	where $K_{qp,\ell}$ is the number of RT rays, and $P_{qp,\ell}^{k}$ and $\tau_{qp,\ell}^{k}$ denote the power and delay of the $k$-th RT ray, respectively. These parameters are directly obtained from the RT simulations described in Section~III-B.
	
	To capture the impact of moving interaction objects, a dynamic component is added on top of the static RT part. Dynamic clusters are introduced to represent groups of MPCs generated by pedestrians and vehicles. Their parameters, including power, delay, angle, and Doppler shift, are extracted from the 5.5~GHz outdoor dynamic measurements by comparing the measured time-varying CIRs with the RT predictions. Then, a GBSM \cite{Wang6GPCM2} is used to simulate the dynamic clusters along the receiver trajectories. The corresponding dynamic CIR between $A_q^T$ and $A_p^R$ at location $\ell$ is given by \cite{QTR_TVT}

\begin{equation}
	\medmath{
		h_{qp,\ell}^{\rm{D}}(t,\tau) \!=\! \sum_{n=1}^{N_{qp,\ell}(t)} \!\! \sum_{n^{\prime}=1}^{M_n} \! \sqrt{P_{qp,\ell}^{n^{\prime}}(t)} \, e^{j2\pi f_c\tau_{qp,\ell}^{n^{\prime}}(t)} \delta(\tau \!-\! \tau_{qp,\ell}^{n^{\prime}}(t))
	}
	\label{h_d}
\end{equation}

	where $N_{qp,\ell}^{(t)}$ is the number of active dynamic clusters at time $t$, and $M_n$ is the number of rays within the $n$-th cluster. The scalar $P_{qp,\ell}^{n'}(t)$ and $\tau_{qp,\ell}^{n'}(t)$ denote the time-varying power and delay of the $n'$-th ray within the $n$-th dynamic cluster. \emph{The detailed extraction and modeling of dynamic clusters, as well as the validation against the outdoor measurements, have been presented in our previous work~\cite{QTR_TVT}}. Here, the same dynamic component is incorporated into the HCM so that the resulting channels can follow the motion of scatterers in real world wireless scenarios. The complete HCM in this stage therefore consists of a static RT component $h_{qp,\ell}^{\mathrm{RT}}(\tau)$, a dynamic component $h_{qp,\ell}^{\mathrm{D}}(t,\tau)$, and a DMC component $h_{qp,\ell}^{\mathrm{DMC}}(\tau)$ that will be introduced next.

	\subsection{DMC Component of Hybrid Channel Model}
	
\begin{figure}[t]
		\centering
		\includegraphics[width=9cm]{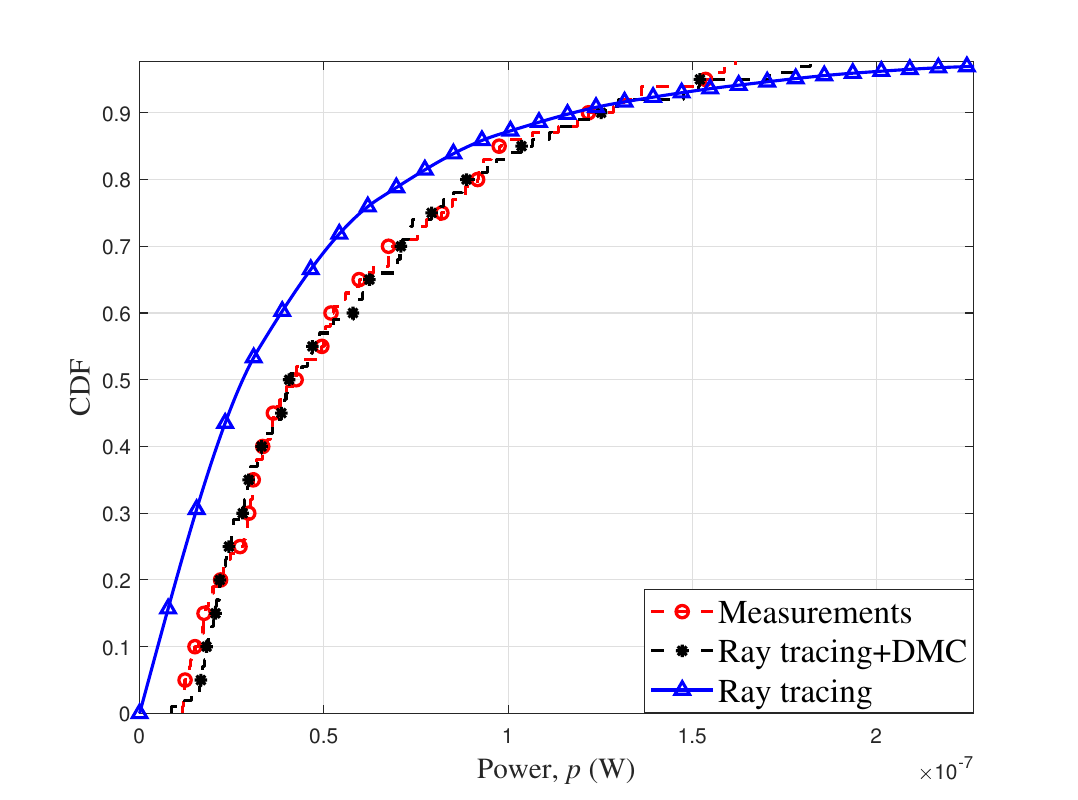}
		\caption{CDFs of MPC power: measurements, pure RT, and RT with DMC.}
		\label{fig:CDFofPower}
	\end{figure}

	From the indoor channel measurements, the DMCs can be extracted by the estimation algorithm proposed in \cite{DMCZhang}. DMC involves more scattered and complex propagation phenomena in wireless transmission, and is an essential part of the wireless propagation channel. It can significantly contribute to the received power across various environments and frequency bands \cite{DMC}. Fig. ~\ref{fig:CDFofPower} compares the cumulative distribution functions (CDFs) of MPC power from the RT simulation and processed measurement data. It is observed that RT simulation fails to fit the measurement results accurately. However, if DMC power is added to the simulated RT MPC power, the combined CDF matches the measurement data well.
	This result indicates that, in indoor scenarios, RT simulations cannot directly replace actual measurements.
	
	 Meanwhile, the result provides guidance for characterizing the differences between RT and measurements by DMC, further generating more accurate channels. Conventional RT methods primarily focused on specular reflections. While these specular paths dominate propagation in many scenarios, they cannot fully explain the diffuse scattering processes that produce DMC \cite{DMC}. The generation of DMC depends significantly on the physical properties of the environment and the operating frequency. Due to this complexity, accurately modeling DMC requires considering multiple scattering mechanisms, often employing statistical descriptions \cite{DMC_AWPL}. The observations in \cite{DMC_LiYuXiao} indicated that the angular and delay power spectra of DMC are not simply random white noise. Instead, they show clear correlations with the SMCs in channel. In~\cite{DMCZhang}, it was suggested that scatterers causing DMC are typically smaller and located near those responsible for SMCs, usually within the same cluster.
	
	In the proposed HCM, DMC is therefore modeled by cluster pairs located around the obstacles along the RT propagation paths, including the first-bounce cluster $C^A_k$ at Tx side and the last-bounce cluster $C^Z_k$ at Rx side. The resulting DMC component $h_{qp,\ell}^{\mathrm{DMC}}(\tau)$ is added to \eqref{h_rt} and \eqref{h_d} to form a more accurate static–dynamic channel representation.
	
	\begin{equation}
		h_{qp,\ell}^{\rm{DMC}}(\tau)=\sum_{k=1}^{K_{qp,\ell}}\sum_{k^{\prime}=1}^{M_{k}}\sqrt{P_{qp,\ell}^{k^{\prime}}}\cdot e^{j2\pi f_{c}\tau_{qp,\ell}^{k^{\prime}}}\cdot\delta(\tau-\tau_{qp,\ell}^{k^{\prime}})
	\end{equation}
	where $P_{qp,\ell}^{k^{\prime}}$ and $\tau_{qp,\ell}^{k^{\prime}}$ are the power and delay of rays $k^{\prime}$ within DMC cluster pairs.
	
	It is worth noting that only the $h_{qp,\ell}^{\rm{D}}(t,\tau)$ is related to time $t$, while the RT and DMC components are only decided by the transceiver location $\ell$. By introducing the DMC and dynamic clusters, the proposed RT-based HCM is capable to provide accurate, temporal channel information in response to changes in the physical environment.

	\section{Data-driven stage of the Channel Map Construction}

	\begin{figure*}[h]
	\centering
	\hspace{-1mm}
	\includegraphics[width=18cm]{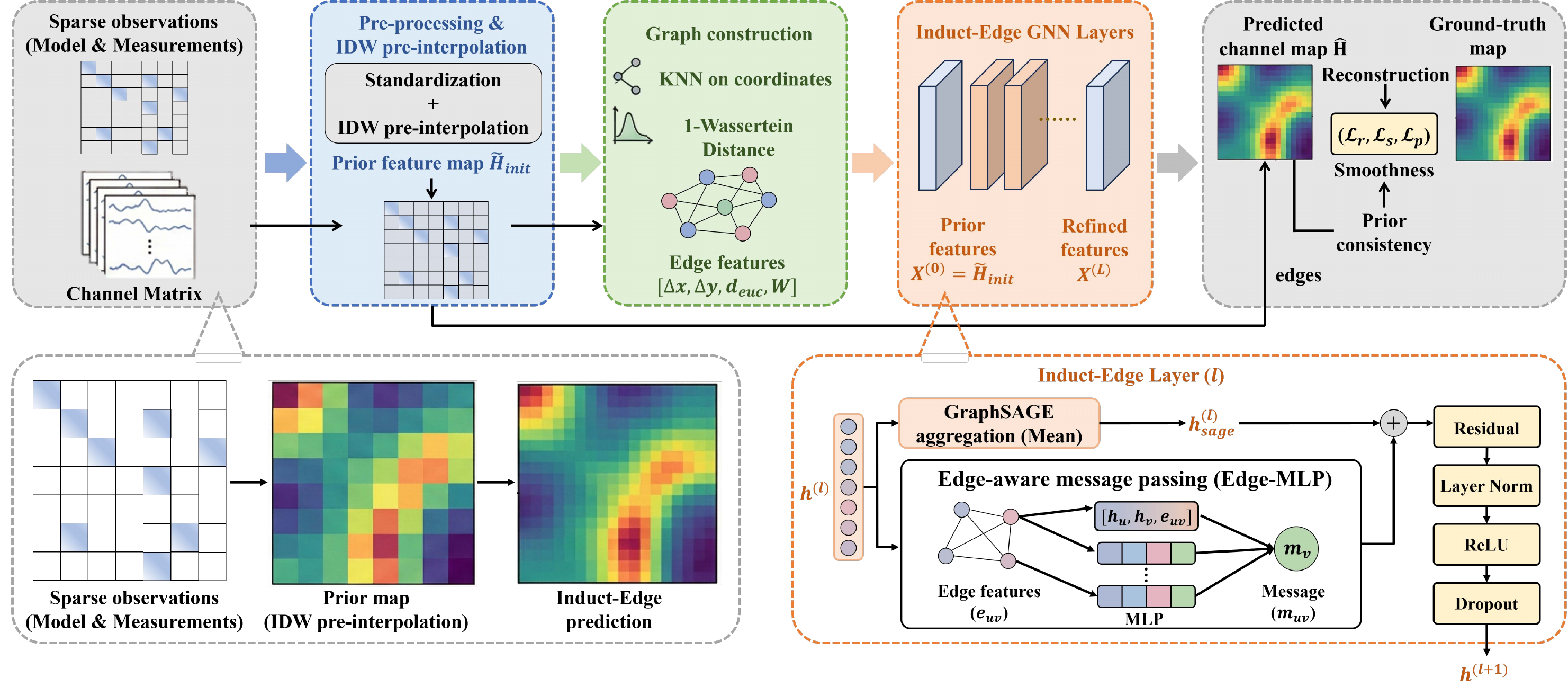}
	\caption{Diagram of the proposed induct-edge graph neural network.}
	\label{fig:DelayPSD_all}
\end{figure*}

	Our methodology addresses the challenge of interpolating discrete and irregular channel data by formulating the problem on a graph. A comprehensive framework centered on a hybrid GNN architecture is proposed and designed to be inductive, scalable, and aware of both spatial geometry and channel-specific statistics. Specifically, the inductive capability enables the DMcD framework to handle dynamic channel variations without the latency of retraining. It allows for timely spatial interpolation whenever the environment changes, thereby maintaining a real-time and accurate perception foundation for agents. The framework is presented in three main steps: (A) the system formulation, feature engineering, and a novel graph construction process, (B) the hybrid inductive GNN architecture for message passing, and (C) the composite training objective and model implementation details.
	
	\subsection{System Formulation and Graph Construction}
	
	At first, a set of $n$ spatial locations are defined on a 2D plane, denoted by their coordinates $\{(x_k, y_k)\}_{i=k}^{n}$. Each location $k$ is modeled as a node in a graph $\mathcal{V}$, and each node is associated with a complex-valued channel matrix $h_k \in \mathbb{C}^{N\times L}$, where where $N$ denotes the number of receiving antennas and $L$ represents the number of subcarriers. In a practical deployment, measurements are discrete, meaning the ground truth channel $h_i$ for a small subset of nodes $i \in \mathcal{V}_{\mathrm{obs}} \subset \mathcal{V}$ is observed. The central objective of our framework is to accurately predict the channel matrices $\hat{\mathbf{h}}_j$ for all unobserved nodes $j \in \mathcal{V}_{\mathrm{unobs}}$, where $\mathcal{V}_{\mathrm{unobs}} = \mathcal{V} \setminus \mathcal{V}_{\mathrm{obs}}$.
	
	As GNNs operate on real-valued feature vectors, a critical first step is the conversion of these complex quantities. The initial node feature $x_k^{(0)}$ is defined using one of two representations: (i) Concatenation, where the real and imaginary parts are flattened and stacked, $x_k^{(0)} = [\text{vec}(\Re(h_k)), \text{vec}(\Im(h_k))] \in \mathbb{R}^{2NL}$, or (ii) Polar Decomposition, $x_k^{(0)} = [\text{vec}(|h_k|), \text{vec}(\angle h_k)] \in \mathbb{R}^{2NL}$. The latter may involve $\sin/\cos$ embeddings of the phase to handle wrapping discontinuities. For all observed nodes $i \in \mathcal{V}_{\mathrm{obs}}$, these features are standardized to have zero mean and unit variance.
	
	A significant challenge arises for unobserved nodes  $j \in \mathcal{V}_{\mathrm{unobs}}$, which lack initial features. To address this and to provide the network with a strong, smooth baseline, we compute an initial estimate $\tilde{\mathbf{h}}_j^{\mathrm{init}}$ for these nodes using Linear Pre-Interpolation. This estimate serves as a "prior" for the GNN to refine. Inverse-distance weighted (IDW)~\cite{IDW} interpolation is employed, where the contribution of observed neighbors is weighted by the inverse of their Euclidean distance $d(j,i)$ \cite{IDW}
\begin{equation}
	\alpha_{ji} = \frac{d(j,i)^{-p}}{\sum_{k\in\mathcal{V}_{\mathrm{obs}}} d(j,k)^{-p}}
\end{equation}
where $p$ is a positive real number known as the power parameter. It controls how quickly the influence of a known data point declines as the distance increases. Therefore, the initial predicted channel $\tilde{\mathbf{h}}_j^{\mathrm{init}}$ is calculated as
\begin{equation}
	\tilde{\mathbf{h}}_j^{\mathrm{init}} = \sum_{i\in\mathcal{V}_{\mathrm{obs}}} \alpha_{ji} \mathbf{h}_i.
\end{equation} 
	This resulting prior (converted to its real-valued form) is then used as the initial feature $x_j^{(0)}$ for all unobserved nodes.
	
		\begin{algorithm}[h]
		\caption{Data-driven stage via Induct-Edge GNN}
		
		\small
		\label{alg:induct_edge_gnn}
		\begin{algorithmic}[1]
			\REQUIRE 
			Spatial locations $\mathcal{P} = \{(x_k, y_k)\}_{k=1}^n$; 
			Observed subset $\mathcal{V}_{\mathrm{obs}}$, Unobserved subset $\mathcal{V}_{\mathrm{unobs}}$;
			Discrete measurements $\mathbf{h}_{\mathrm{obs}} = \{\mathbf{h}_i \mid i \in \mathcal{V}_{\mathrm{obs}}\}$;
			Hyperparameters: layers $S$, smoothness weight $\lambda$, prior weight $\eta$, learning rate $\gamma$.
			\ENSURE Predicted Channel Maps $\hat{\mathbf{h}} = \{\hat{\mathbf{h}}_k \mid \forall k \in \mathcal{V}\}$
			
			\STATE \textbf{Phase 1: Graph Construction \& Feature Engineering}
			\STATE \COMMENT{Construct W-KNN Graph based on distribution similarity}
			\STATE Compute 1-Wasserstein distance $W(u,v)$ between channel distributions of node pairs;
			\STATE Build graph $\mathcal{G}=(\mathcal{V}, \mathcal{E})$ connecting K-nearest neighbors via $W(u,v)$;
			\STATE Compute edge features $e_{uv} = [\Delta x_{uv}, \Delta y_{uv}, d_{uv}^{\mathrm{Euc}}, W(u,v), \mathrm{LoS/NLoS}_{uv}] \in \mathbb{R}^{D_e}$;
			\STATE Compute edge weights $w_{uv} = \exp(-W(u,v)/\tau)$ for smoothness regularization;
			
			\STATE \textbf{Phase 2: Pre-Interpolation (Prior Generation)}
			\FOR{each $j \in \mathcal{V}_{\mathrm{unobs}}$} 
			\STATE \COMMENT{Focus on Unobserved nodes}
			\STATE Calculate weights $\alpha_{ji} = d(j,i)^{-p} / \sum_{m \in \mathcal{V}_{\mathrm{obs}}} d(j,m)^{-p}$ (IDW);
			\STATE Generate prior $\tilde{\mathbf{h}}_j^{\mathrm{init}} \leftarrow \sum_{i \in \mathcal{V}_{\mathrm{obs}}} \alpha_{ji} \mathbf{h}_i$;
			\ENDFOR
			\STATE Initialize node features $x_k^{(0)}$ for all $k \in \mathcal{V}$ using Concatenation; 
			
			\STATE \textbf{Phase 3: Inductive Training (Hybrid Message Pass)}
			\FOR{epoch $t = 1$ to $T$}
			\FOR{each mini-batch $\mathcal{B} \subseteq \mathcal{V}$}
			\STATE Initialize hidden states $z_k^{(0)} \leftarrow x_k^{(0)}$ for all $k \in \mathcal{B} \cup \mathcal{N}(\mathcal{B})$;
			\FOR{layer $l = 1$ to $S$}
			\FOR{each node $k \in \mathcal{B}$} 
			\STATE \COMMENT{\textit{Step A: GraphSAGE Generalized Aggregation}}
			\STATE Sample neighbors $\mathcal{N}^{(l)}(k)$ from graph $\mathcal{G}$;
			\STATE $m_k^{(l)} \leftarrow \text{MEAN}(\{z_u^{(l-1)} \mid \forall u \in \mathcal{N}^{(l)}(k)\})$;
			\STATE Intermediate: $\tilde{z}_k^{(l)} \leftarrow \sigma(\mathbf{W}_{\mathrm{agg}}^{(l)} \cdot [z_k^{(l-1)} \Vert m_k^{(l)}] + \mathbf{b}^{(l)})$;
			\STATE Apply LayerNorm and Dropout on $\tilde{z}_k^{(l)}$;
			
			\STATE \COMMENT{\textit{Step B: Edge-Aware Refinement (ECC)}}
			\STATE Generate dynamic filter: $\mathbf{\Theta}_{\phi}(e_{ku}) \leftarrow \text{MLP}_{\phi}(e_{ku})$;
			\STATE Refine: $z_k^{(l)} \leftarrow \sigma\left(\sum_{u \in \mathcal{N}^{(l)}(k)} \mathbf{\Theta}_{\phi}(e_{ku})\tilde{z}_u^{(l)} + \mathbf{W}_{\mathrm{self}}^{(l)}\tilde{z}_k^{(l)}\right)$;
			\STATE Residual Connection: $z_k^{(l)} \leftarrow z_k^{(l)} + z_k^{(l-1)}$;
			\ENDFOR
			\ENDFOR
			
			\STATE \COMMENT{\textit{Output \& Composite Loss Calculation}}
			\STATE Prediction: $y_k \leftarrow \text{MLP}_{\mathrm{out}}(z_k^{(S)})$, reconstructed to $\hat{\mathbf{h}}_k$;
			\STATE $\mathcal{L}_r \leftarrow \frac{1}{|\mathcal{V}_{\mathrm{obs}}|}\sum_{i\in\mathcal{V}_{\mathrm{obs}}} \frac{\|\hat{\mathbf{h}}_i - \mathbf{h}_i\|_2^2}{\|\mathbf{h}_i\|_F^2 + \epsilon}$ \COMMENT{Reconstruction (Observed $i$)};
			\STATE $\mathcal{L}_s \leftarrow \frac{\lambda}{|\mathcal{E}|}\sum_{(i,j)\in \mathcal{E}} w_{ij} \|\hat{\mathbf{h}}_i - \hat{\mathbf{h}}_j\|_F^2 = \frac{\lambda}{|\mathcal{E}|} \mathrm{tr}\big( \hat{\mathbf{h}}^\top L \hat{\mathbf{h}}\big)$\COMMENT{Smoothness (Edges $u,v$)};
			\STATE $ \mathcal{L}_p \leftarrow \frac{\eta}{|\mathcal{V}_{\mathrm{unobs}}|}\sum_{i\notin\mathcal{V}_{\mathrm{obs}}} \|\hat{\mathbf{h}}_j - \tilde{\mathbf{h}}_j^{\mathrm{init}}\|_F^2 $\COMMENT{Prior Consistency (Unobserved $j$)};
			\STATE Backpropagation: $\theta \leftarrow \text{AdamW}(\nabla (\mathcal{L}_r + \mathcal{L}_s + \mathcal{L}_p))$;
			\ENDFOR
			\ENDFOR
			
			\STATE \textbf{Phase 4: Inference}
			\STATE $\hat{\mathbf{h}} = \text{ForwardPass}(\mathcal{G}, \mathbf{X}^{(0)})$ with trained weights $\theta^*$.
			
		\end{algorithmic}
		
	\end{algorithm}

	The graph topology $G=(\mathcal{V}, \mathcal{E})$ is paramount. Relying on naive Euclidean distance to define edges is a flawed heuristic, as it fails to capture the underlying propagation physics. To overcome this, a K-Nearest Neighbors (KNN) graph is constructed based on the Wasserstein distance. This metric quantifies the similarity between the distributions of channel characteristics. For each node $k$, an empirical distribution $\mu_k$ is formed from the magnitudes of its $N \times L$ channel entries. The Wasserstein distance $W(p,q)$ between two nodes $p$ and $q$ measures the minimum “cost” to transform $\mu_p$ into $\mu_q$ \cite{channelGAN}
	\begin{equation}
	W(p,q) = \inf_{\gamma\in\Gamma(\mu_p,\mu_q)} \int_{\mathbb{R}\times\mathbb{R}} |u-v| d\gamma(u,v).
	\end{equation}
	
	Since the channel magnitudes form a univariate distribution, this calculation simplifies to the computationally efficient $L_1$ distance between the CDFs, $W(p,q) = \int_{\mathbb{R}} |F_p(z)-F_q(z)| dz$. This allows the graph to connect nodes that share similar MPC fading profiles, even if they are not geometrically adjacent.
	
	Finally, for each edge $(u,v) \in \mathcal{E}$ in the resulting W-KNN graph, a rich, multi-dimensional edge feature vector $e_{uv}$ is computed. This vector is crucial for the edge-aware refinement stage and is defined as
	\begin{equation}e_{uv} = [\Delta x_{uv}, \Delta y_{uv}, d_{uv}^{\mathrm{Euc}}, W(u,v), \mathrm{LoS/NLoS}_{uv}]\end{equation}
	where $\Delta x_{uv} = x_u - x_v$ and $\Delta y_{uv} = y_u - y_v$ denotes the relative displacement in the x-coordinate and y-coordinate, respectively. $d_{uv}^{\mathrm{Euc}} = \sqrt{\Delta x_{uv}^2 + \Delta y_{uv}^2}$, denoting the Euclidean distance between the nodes. $\mathrm{LoS/NLoS}_{uv}$ is a binary or categorical indicator representing whether the path between $u$ and $v$ is likely LoS or NLoS.
	
	Edge weights $w_{uv}=\exp(-W(u,v)/\tau)$ are also computed to form a symmetric adjacency matrix $\tilde{A}$, which is used in the smoothness regularizer.

	\subsection{Inductive GNN Architecture and Message Passing}
	
	Our GNN architecture is designed to be both scalable and inductive, meaning it can generalize to new, unseen nodes and measurements without requiring complete retraining. To achieve this, we adopt the hierarchical neighbor sampling strategy from graph sample and aggregate (GraphSAGE) network \cite{GraphSAGE_1}. Instead of processing the entire graph, training is performed on mini-batches. For $B$ seed nodes, we perform $S$ hops of message passing. At each layer $l$, each node samples a fixed size, $k_l$, of neighbors from its 1-hop neighborhood $\mathcal{N}^{(l)}(k)$. This approach ensures that the computational cost per iteration remains controlled and independent of the full graph size. The core of our model is a hybrid message-passing block that is iterated $S$ times. Let $z_k^{(l)} \in \mathbb{R}^{D_l}$ denote the hidden feature of node $k$ at layer $l$. Each block is a two-step process: a general-purpose aggregation followed by an edge-aware refinement.
	
	\subsubsection{Generalized Aggregation} First, a standard GraphSAGE update is performed. This step learns a generalized, inductive aggregation function by computing the mean of the sampled neighbors' features \cite{GraphSAGE_1}
	\begin{equation}m_k^{(l)} = \frac{1}{|\mathcal{N}^{(l)}(k)|}\sum_{u\in\mathcal{N}^{(l)}(k)} z_u^{(l)}.\end{equation}
	This aggregated message $m_k^{(l)}$ is concatenated with the node's current self-representation $z_k^{(l)}$. The result $[z_k^{(l)} \Vert m_k^{(l)}]$ is passed through a learnable offine transformation with a non-linearity (e.g., ReLU or modReLU) to produce an intermediate representation $\tilde{z}_k^{(l+1)}$, which is calculated as \cite{GraphSAGE_1}
	\begin{equation}\tilde{z}_k^{(l+1)} = \sigma\big(\mathbf{W}_{\mathrm{agg}}^{(l)} \cdot [z_k^{(l)} \Vert m_k^{(l)}] + \mathbf{b}^{(l)}\big)\end{equation}
	where $\mathbf{W}_{\mathrm{agg}}^{(l)} \in \mathbb{R}^{D_{l+1} \times 2D_l}$ and $\mathbf{b}^{(l)} \in \mathbb{R}^{D_{l+1}}$ denote the learnable weight matrix and bias vector for layer $l$, respectively. This intermediate representation is then stabilized using Layer Normalization and dropout.
	
	\subsubsection{Edge-Aware Refinement} Second, this generalized feature $\tilde{z}_k^{(l+1)}$ is refined using Edge-Conditioned Convolution (ECC). This step is critical as it leverages the rich edge features $e_{ku}$ previously engineered. ECC employs a small filter-generating network (implemented as Multi-Layer Perceptron, $\mathrm{MLP}_{\phi}$) that dynamically computes a specific transformation matrix $\mathbf{\Theta}_{\phi}(e_{ku}) \in \mathbb{R}^{D_{l+1}\times D_{l+1}}$ for each unique edge instance
	\begin{equation}\mathbf{\Theta}_{\phi}(e_{ku}) = \mathbf{MLP}_{\phi}(e_{ku}).\end{equation}
	
	The final hidden state $z_k^{(l+1)}$ is then computed by applying these edge-specific filters to the messages from the neighborhood, allowing the model to learn complex, non-linear dependencies driven by physical channel correlations
	\begin{equation}z_k^{(l+1)} = \sigma\Big(\sum_{u\in\mathcal{N}^{(l)}(k)} \mathbf{\Theta}_{\phi}(e_{ku}) \tilde{z}_u^{(l+1)} + \mathbf{W}_{\mathrm{self}}^{(l)}\tilde{z}_k^{(l+1)}\Big)\end{equation}
	where $\mathbf{W}_{\mathrm{self}}^{(l)} \in \mathbb{R}^{D_{l+1} \times D_{l+1}}$ represents the learnable weight matrix for the self-loop connection, ensuring the node retains its own contextual information during the refinement. This mechanism allows the network to, for instance, learn to weight messages from LoS neighbors or adapt its aggregation based on the Wasserstein similarity. A residual path  ($z_k^{(l+1)} \leftarrow z_k^{(l)} + z_k^{(l+1)}$) is added to facilitate the training of deeper GNNs.
	
	After $S$ message-passing blocks, the final node embedding $z_k^{(S)}$ encapsulates rich contextual information from its $S$-hop neighborhood. This embedding is passed through a final prediction head $\mathrm{MLP}_{\mathrm{out}}$ to map to the target dimension $\mathbb{R}^{2NL}$
	\begin{equation}
	y_k = \mathrm{MLP}_{\mathrm{out}}(z_k^{(S)}).
	\end{equation}
	
	The output vector $y_k \in \mathbb{R}^{2NL}$ is split into two halves. The first $NL$ elements form the real part (or magnitude), and the remaining $NL$ elements form the imaginary part (or phase). For the concatenation case, the reconstruction is
	\begin{equation}\hat{\mathbf{h}}_k = \mathrm{reshape}(y_k^{[1:NL]}) + j \cdot \mathrm{reshape}(y_k^{[NL+1:2NL]}).\end{equation}
	
	\vspace{-1em}
	\subsection{ Composite Training Objective and Implementation}
	
	To ensure the model produces predictions that are not only accurate but also physically plausible and robust in sparse regions, we design a composite loss function $\mathcal{L} = \mathcal{L}_r + \mathcal{L}_s + \mathcal{L}_p$. This objective balances data fidelity with two powerful graph-aware regularization terms.
	
	The primary term is the reconstruction loss $\mathcal{L}_r$. This loss is enforced only on the observed nodes $i \in \mathcal{V}_{\mathrm{obs}}$ where ground truth is available. We employ a Normalized Mean Squared Error (NMSE)-weighted MSE, which normalizes the error by the power of the ground truth channel. This prevents high-power (e.g., LoS) locations from dominating the loss signal
	\begin{equation}
	\mathcal{L}_r = \frac{1}{|\mathcal{V}_{\mathrm{obs}}|}\sum_{i\in\mathcal{V}_{\mathrm{obs}}} \frac{\|\hat{\mathbf{h}}_i - \mathbf{h}_i\|_2^2}{\|\mathbf{h}_i\|_F^2 + \epsilon}
	\end{equation}
	where $\|\cdot\|_F$ denotes the Frobenius norm.
	
	The second term is an edge-aware smoothness loss $\mathcal{L}_s$, which acts as a graph Laplacian regularizer. This term encourages the predictions $\hat{\mathbf{h}}_i$ and $\hat{\mathbf{h}}_j$ to be similar if their corresponding nodes $(i,j)$ are connected by a high-weight edge $w_{ij}$ (i.e., they have high channel similarity). This promotes spatially coherent predictions that respect the W-distance manifold
	\begin{equation}
	\mathcal{L}_s = \frac{\lambda}{|\mathcal{E}|}\sum_{(i,j)\in \mathcal{E}} w_{ij} \|\hat{\mathbf{h}}_i - \hat{\mathbf{h}}_j\|_F^2 = \frac{\lambda}{|\mathcal{E}|} \mathrm{tr}\big( \hat{\mathbf{h}}^\top L \hat{\mathbf{h}}\big)
	\end{equation}
	where $L=D-\tilde{A}$ is graph Laplacian. $\lambda$ is a tuning parameter.
	
	Finally, we introduce a prior consistency loss $\mathcal{L}_p$. This novel regularizer is enforced only on the unobserved nodes $j \in \mathcal{V}_{\mathrm{unobs}}$. It penalizes the model's prediction $\hat{\mathbf{h}}_i$ for deviating from the smooth, pre-interpolated prior $\tilde{\mathbf{h}}_i^{\mathrm{init}}$. This is crucial for stabilizing the network's behavior and preventing unrealistic artifacts in large regions with no measurement data
	\begin{equation}
	\mathcal{L}_p = \frac{\eta}{|\mathcal{V}_{\mathrm{unobs}}|}\sum_{i\notin\mathcal{V}_{\mathrm{obs}}} \|\hat{\mathbf{h}}_j - \tilde{\mathbf{h}}_j^{\mathrm{init}}\|_F^2
	\end{equation}
	where $\eta$ is a tuning hyperparameter.
	
	The model is trained end-to-end using the AdamW optimizer with a cosine decay learning rate schedule and gradient clipping. The inductive nature of the framework is a key practical advantage: when new measurements arrive, updated map predictions can be generated via a single forward pass on the updated graph, further avoiding the retraining.

\section{Experiment Results and Evaluation}

\begin{table}[!t]
	\centering
	\small 
	\caption{HYPERPARAMETER SETTINGS OF THE INDUCT-EDGE GNN.}
	\setlength{\tabcolsep}{3mm}
	\label{tab:hyper_inductedge}
	\arrayrulecolor{black!70} 
	\renewcommand{\arraystretch}{1.4}
	\setlength{\arrayrulewidth}{0.8pt} 
	\begin{tabular}{c|c} 
		\hline
		\hline
		\rowcolor{gray!15} \textbf{Parameters} & \textbf{Value} \\ \hline
		SAGEConv layers of encoder & 2 \\ 
		\rowcolor{gray!15} 	Number of GNN layers & 2 \\ 
		Hidden channels & 64 \\ 
		\rowcolor{gray!15} 	Dropout rate & 0.1 \\
		Activation function & ReLU \\
		\rowcolor{gray!15} Learning rate ($\gamma$) & $10^{-3}$ \\
		Optimizer & AdamW (wd $10^{-4}$) \\
		\rowcolor{gray!15} Batch size & 1 (full graph) \\
		Maximum epochs ($T$) & 500 \\
		\rowcolor{gray!15} Reconstruction loss ($L_r$) & NMSE on observed nodes \\
		Smoothness loss ($L_s$) & edge-weighted MSE on graph \\
		\rowcolor{gray!15} Prior consistency loss ($L_p$) & MSE on unobserved nodes \\
		Smoothness weight ($\lambda$) & 0.1 \\
		\rowcolor{gray!15} Prior weight ($\eta$) & 0.1 \\
		IDW power parameter & 2 \\
		\rowcolor{gray!15} $k$-NN neighbors in graph & 10 \\
		\hline
		\hline
	\end{tabular}
\vspace{-2em}
\end{table}

\subsection{Evaluation Setup}

This section evaluates the proposed DMcD framework from three levels: the model-driven stage (HCM), the data-driven stage (InductE-GNN), and the overall DMcD channel map construction framework.
Measurements in the indoor scenario with ultra-dense sampling and rich DMCs and the outdoor scenario with moving objects are considered as baseline.

\textbf{Data sources and usage:}
For the model-driven stage, the RT simulator provides specular static paths.
The indoor measurements are mainly used to calibrate the DMC-related statistics in the HCM, while the outdoor measurements are used to extract and validate the parameters of dynamic clusters.
For the data-driven stage, a 2D spatial grid is formed by uniformly sampled receiver locations.
At each grid node, the channel is represented by the measured/simulated channel matrix.
Only a subset of nodes are treated as observed, and the remaining nodes are regarded as unobserved targets for interpolation. The detailed hyperparameter setting is shown in Table.~\ref{tab:hyper_inductedge}.

\textbf{Sampling protocol:}
Observed nodes are selected with approximately uniform spacing over the grid to emulate sparse probing. The sampling density is swept by changing the number of observed nodes, which enables an assessment of the practical measurement requirement for reliable interpolation. Unless otherwise stated, the same sampling rule is used for all learning-based baselines to ensure a consistent comparison.

\textbf{Metrics:}
For HCM validation, the statistical dispersion characteristics are compared with measurements, including the CDFs of root-mean-square (RMS) delay spread (DS) $\tau_{\mathrm{rms},\ell}$ and RMS angular spread (AS) $\sigma_{\mathrm{rms},\ell}$ at each location $\ell$, as well as representative delay PSD and angle-domain spectra.
The calculation of these quantities follows the definitions in~\cite{RMSDS}
\begin{equation}
	\tau^{\rm{rms}}_{\ell}=\sqrt{\frac{\sum_{k=1}^\mathcal{K}(\tau^k_{\ell}(t)-\tau_{\ell}^{\rm{mean}}(t))^2|h^k_{\ell}(t))|^2}{\sum_{k=1}^\mathcal{K}|h^k_{\ell}(t))|^2}}
\end{equation}

\begin{equation}
	\sigma^{\rm{rms}}_{\ell}=\sqrt{\frac{\sum_{k=1}^\mathcal{K}(\phi^k_{\ell}(t)-\phi_{\ell}^{\rm{mean}}(t))^2|h^k_{\ell}(t))|^2}{\sum_{k=1}^\mathcal{K}|h^k_{\ell}(t))|^2}}
\end{equation}
where $\tau^{\rm{mean}}_{\ell}$ and $\sigma^{\rm{mean}}_{\ell} $ denote the mean excess delay and angle at location $\ell$, respectively. They can be calculated as  
\begin{equation}
	\tau^{\rm{mean}}_{\ell}=\frac{\sum_{k=1}^\mathcal{K}\tau^k_{\ell}(t)|h^k_{\ell}(t))|^2}{\sum_{k=1}^\mathcal{K}|h^k_{\ell}(t))|^2}
\end{equation}
\begin{equation}
	\sigma^{\rm{mean}}_{\ell}=\frac{\sum_{k=1}^\mathcal{K}\phi^k_{\ell}(t)|h^k_{\ell}(t))|^2}{\sum_{k=1}^\mathcal{K}|h^k_{\ell}(t))|^2}.
\end{equation}

For InductE-GNN evaluation, the main metric is the NMSE on unobserved nodes, which can be calculated as
\begin{equation}
	\mathrm{NMSE}=\frac{\sum_{i\in \mathcal{V}_{\rm{unobs}}}\|\hat{\mathbf{h}}_i-\mathbf{h}_i\|_2^2}{\sum_{i\in \mathcal{V}_{\rm{unobs}}}\|\mathbf{h}_i\|_2^2}
	\label{eq:nmse_inducte}
\end{equation}
where $\mathbf{h}_i$ and $\hat{\mathbf{h}}_i$ are the ground-truth and predicted feature vectors, respectively.
For overall DMcD evaluation, spatial channel maps of received power, RMS delay spread, and RMS angular spread are compared against measurement-based references.
In addition, the temporal consistency is evaluated in the outdoor dynamic scenario by comparing time-varying delay PSD patterns along a trajectory.

\textbf{Baselines and training budget:}
For HCM, the proposed model is compared with RT-only results and measurement references.
For data-driven interpolation, GRU, CGAN, and GCN are considered as learning baselines, and IDW is used as a conventional interpolation baseline.
All learning models are tuned under the same training budget, and regularization techniques such as early stopping, dropout, and weight decay are applied when needed to reduce overfitting.

\vspace{-0.5em}

\subsection{Validation of the Proposed Hybrid Channel Model}

\begin{figure*}[t]
	\centering
	\subfloat[]
	{
		\hspace{-5mm}
		\includegraphics[width=5cm]{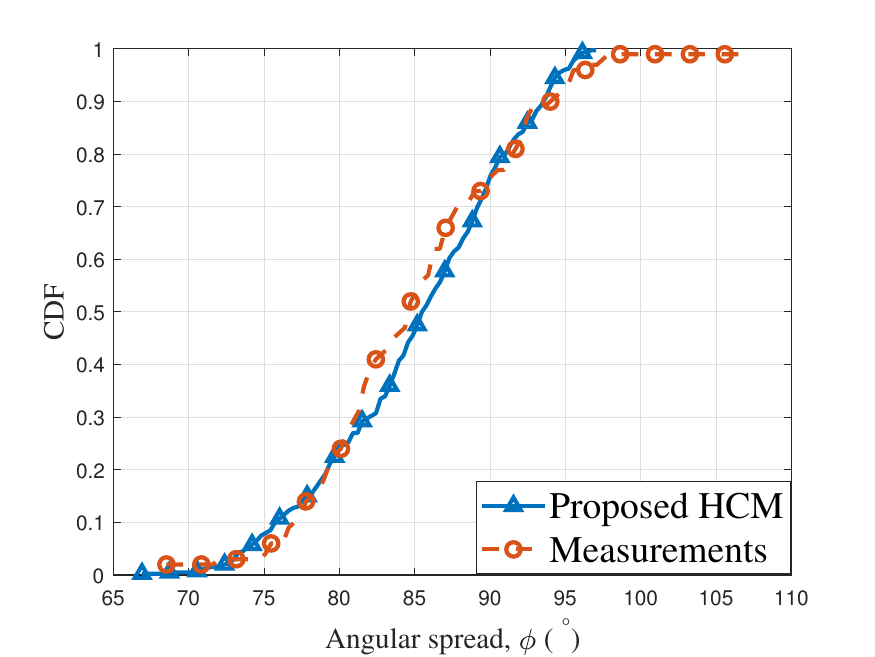}
		
		\label{fig:CDF of AS}
	}
	\subfloat[]
	{
		\hspace{-8mm}
		\includegraphics[width=5cm]{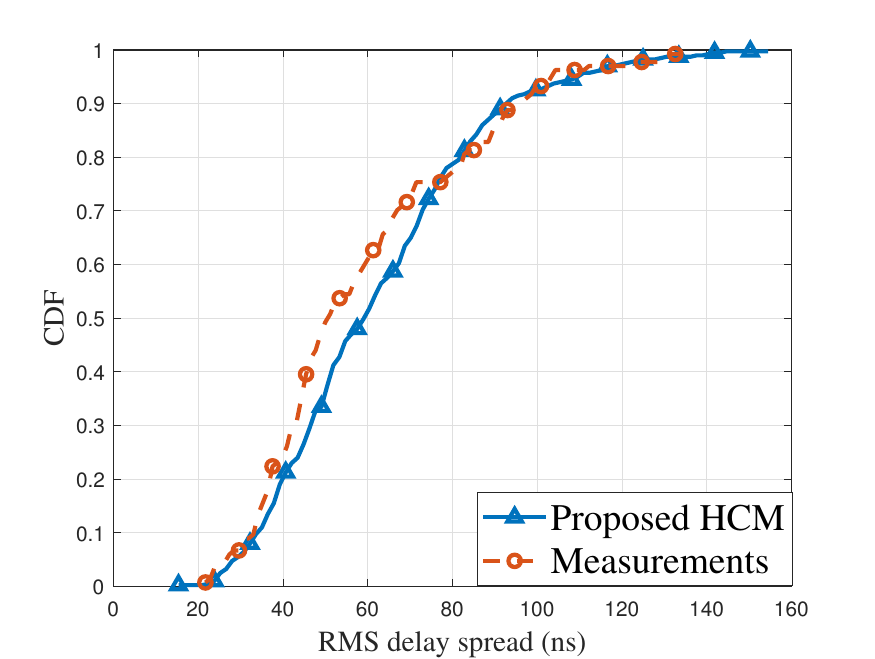}
		\label{fig:CDF of DS}
	}
	\subfloat[]
	{
		\hspace{-8mm}
		\includegraphics[width=5cm]{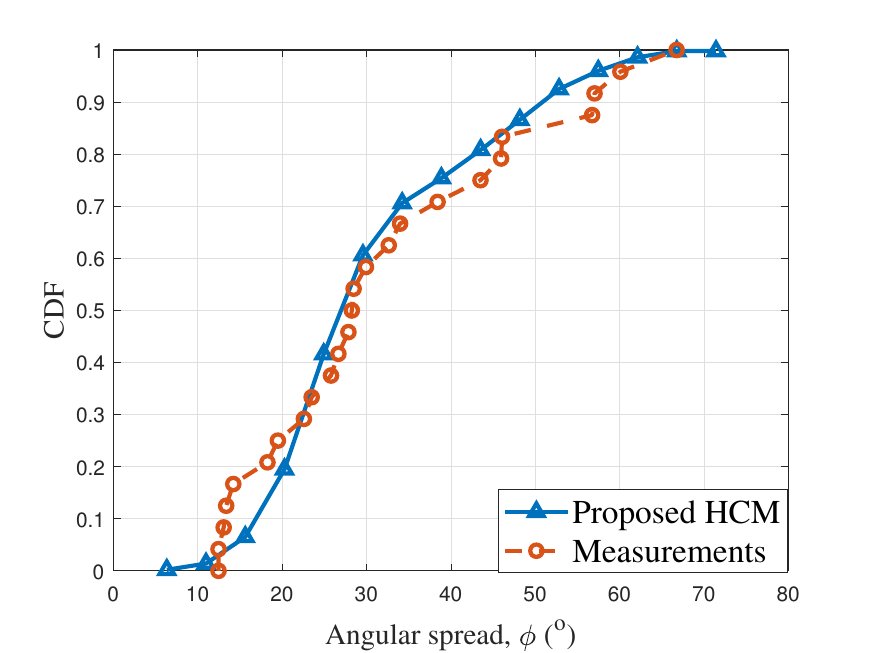}
		\label{fig:CDF of AS out}
	}
	\subfloat[]
	{
		\hspace{-8mm}
		\includegraphics[width=5cm]{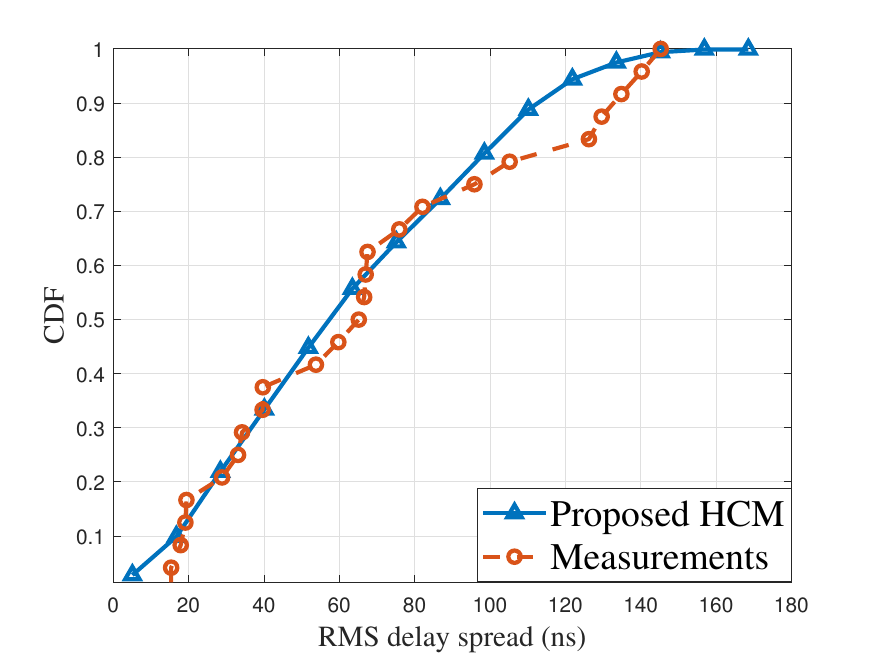}
		\label{fig:CDF of DS out}
	}
	\caption{CDFs of (a) AS (indoor), (b) DS (indoor), (c) AS (outdoor), and (d) DS (outdoor), comparing the proposed HCM with measurements.}
	\vspace{-1em}
	\label{fig:CDF of AS&DS}
\end{figure*}

\begin{figure}[!b]
	\centering
	\includegraphics[width=9cm]{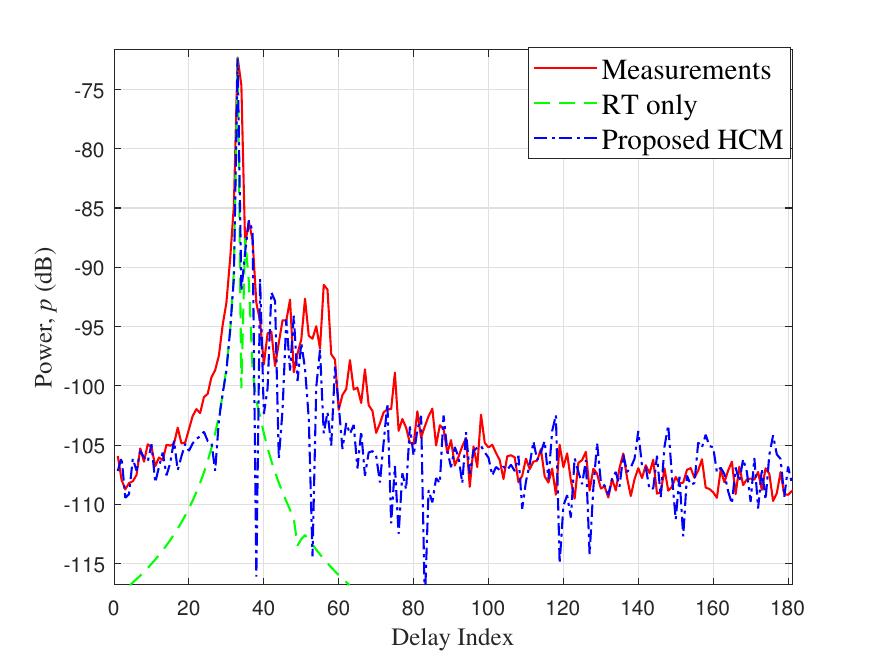}
	\caption{Delay PSDs of measurements, pure RT, and the proposed HCM.}
	\vspace{-1em}
	\label{fig:DelayPSD}
\end{figure}

The accuracy of the proposed RT-based HCM is first evaluated by comparing its statistical characteristics with those obtained from measurements.

Fig.~\ref{fig:CDF of AS&DS} compares the CDFs of RMS angular spread and RMS delay spread between the proposed HCM and the measurements for both indoor and outdoor scenarios. Figs.~\ref{fig:CDF of AS&DS}\subref{fig:CDF of AS} and \subref{fig:CDF of DS} correspond to the indoor 28~GHz campaign, while Figs.~\ref{fig:CDF of AS&DS}\subref{fig:CDF of AS out} and \subref{fig:CDF of DS out} correspond to the outdoor 5.5~GHz campaign. In all cases, the CDFs generated by the proposed HCM agree well with the measured ones, which indicates that the large-scale dispersion characteristics in both delay and angle domains are properly captured. The agreement is especially good in the indoor scenario, where the proposed HCM benefits from the explicit modeling of DMCs. More detailed validation of the dynamic component has been reported in~\cite{QTR_TVT}, and only representative results are shown here.

To highlight the role of DMCs in indoor environments, Fig.~\ref{fig:DelayPSD} further compares the proposed HCM with pure RT. Fig.~\ref{fig:DelayPSD} shows the delay PSDs at a representative indoor location. The RT-only curve can reproduce the LoS path and a few dominant multipaths but clearly underestimates the received power at larger delays. After adding the DMC component, the HCM curve matches the measurements over the whole delay range, especially for delays greater than about 40~ns.

\vspace{-0.5em}
\subsection{Evaluation of the Proposed InductE-GNN}

This subsection evaluates the proposed InductE-GNN for channel map interpolation. 
The task is formulated as a spatial interpolation problem on a $5\times 800$ grid, where each node is associated with a channel matrix and only a subset of nodes are observed. 
The prediction accuracy is measured by the NMSE on the unobserved nodes.
\begin{figure}[t]
	\centering

	\includegraphics[width=9cm]{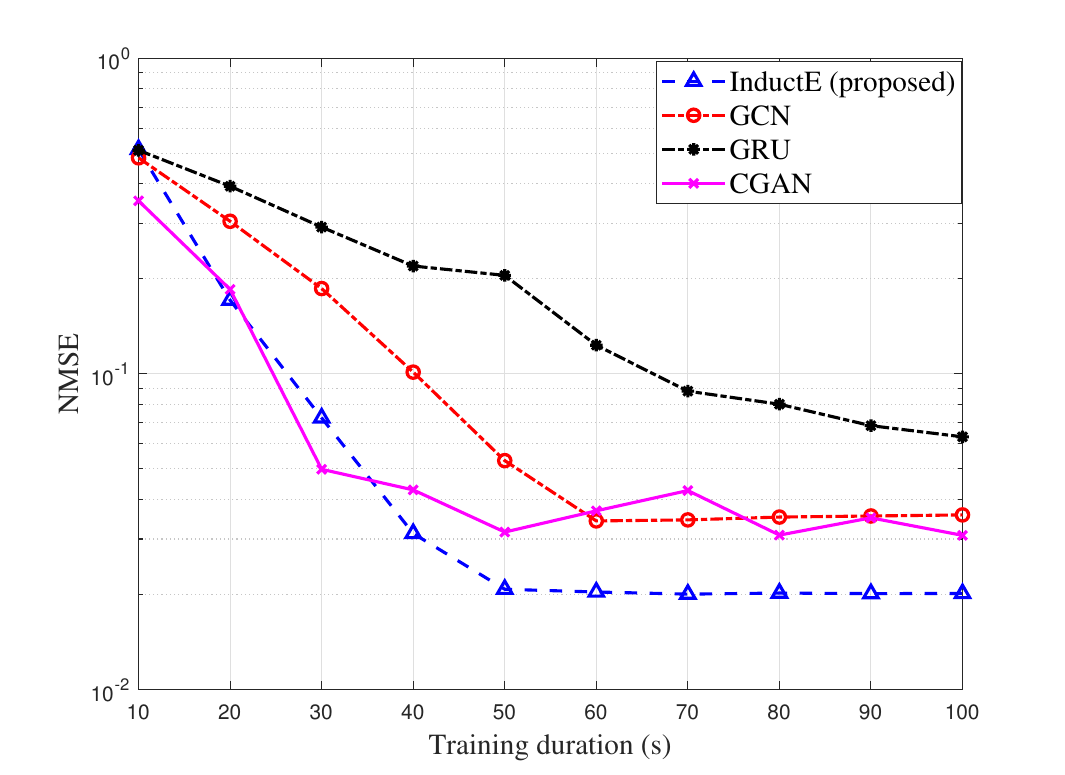}
	\caption{NMSE along training duration for different data-driven methods.}
	\vspace{-1em}
	\label{fig:inducte_time}
\end{figure}

\subsubsection{Comparison with existing data-driven methods}

Fig.~\ref{fig:inducte_time} shows the NMSE versus training duration for GRU, CGAN, GCN, and the proposed InductE-GNN. 
InductE-GNN converges to the lowest NMSE and keeps a clear margin over all baselines, demonstrating its stronger capability for channel map interpolation under sparse observations.

To ensure a fair comparison, all models are carefully tuned under the same time budget. 
For GRU and CGAN, overfitting is mitigated by early stopping (validation NMSE), dropout, and weight decay. 
For CGAN, small instance noise is further injected into the discriminator inputs to stabilize adversarial training and improve generalization.

In Fig.~\ref{fig:inducte_time}, GRU is designed for sequential dependency and thus cannot explicitly exploit the spatial topology, leading to a higher NMSE floor. 
GCN decreases NMSE steadily but saturates early due to isotropic neighborhood averaging: repeated convolutions smooth node features, which denoises the prior but also suppresses local variations needed for accurate recovery at specific locations. 
CGAN reaches a similar NMSE floor to GCN, while showing faster early improvement and mild fluctuations later. This is typical for adversarial training: the generator first learns coarse channel matrix structures via reconstruction loss, but later must balance distribution matching and pointwise NMSE minimization. Since these objectives are not perfectly aligned for deterministic interpolation, a performance floor and small oscillations appear.
In contrast, InductE-GNN achieves a lower NMSE floor because it is designed for inductive spatial interpolation with edge awareness. 
Aggregation enables generalization to unseen nodes and efficient propagation from observed locations, while edge-conditioned message passing leverages informative edge attributes to avoid simple averaging. 
The designs jointly explain the convergence and the lowest NMSE of InductE-GNN.

\subsubsection{Accuracy under different sampling densities and ablation study}
\begin{figure}[t]
	\centering
	\includegraphics[width=9cm]{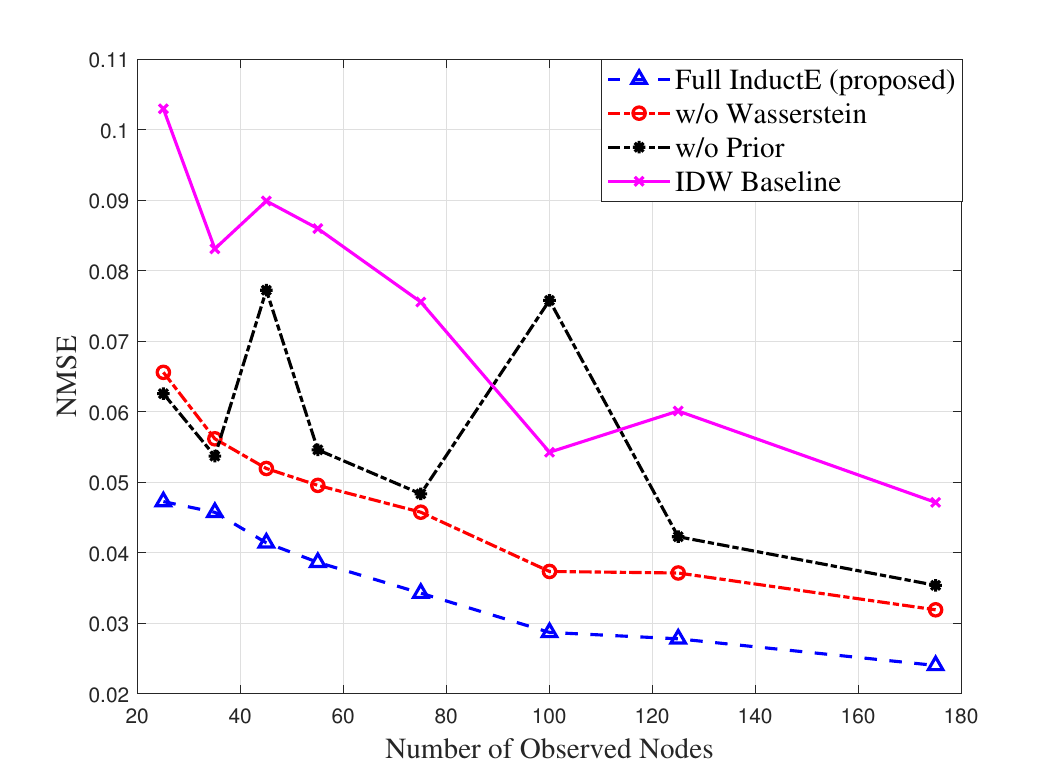}
	\caption{NMSE under different numbers of observed nodes (with ablations).}
	\vspace{-1em}
	\label{fig:inducte_density}
	
\end{figure}
Fig.~\ref{fig:inducte_density} further evaluates InductE-GNN under different numbers of observed nodes. 
As the number of observed nodes increases, the NMSE decreases for all methods, showing that sparse probing is the main bottleneck. 
In particular, the proposed InductE-GNN consistently outperforms the IDW baseline, and the gain is more evident in the sparse regime, which indicates that InductE-GNN can better exploit spatial correlations to predict the channel at a previously unobserved location. Fig.~\ref{fig:inducte_density} also reveals the sampling boundary required for reliable interpolation. Take $\mathrm{NMSE}<0.03$ as an acceptable threshold, a sampling interval of 1.2 m (100 observed nodes) is approximately sufficient for accurate channel map construction, and additional observations bring only marginal gains.

Moreover, Fig.~\ref{fig:inducte_density} provides an ablation study to quantify the contribution of key components. 
Removing the Wasserstein-based edge construction degrades the performance across densities, confirming that similarity-informed edges improve message passing beyond purely geometric neighbors. 
Removing the IDW prior leads to noticeably higher NMSE and less stable trends, suggesting that a smooth physical prior is important for training when observations are limited. 
Overall, the results verify that both the Wasserstein-informed edges and the IDW prior are effective components in the proposed InductE-GNN.

\begin{figure*}[t]
	\centering
	\hspace{-0.8cm}
	\includegraphics[width=18.5cm]{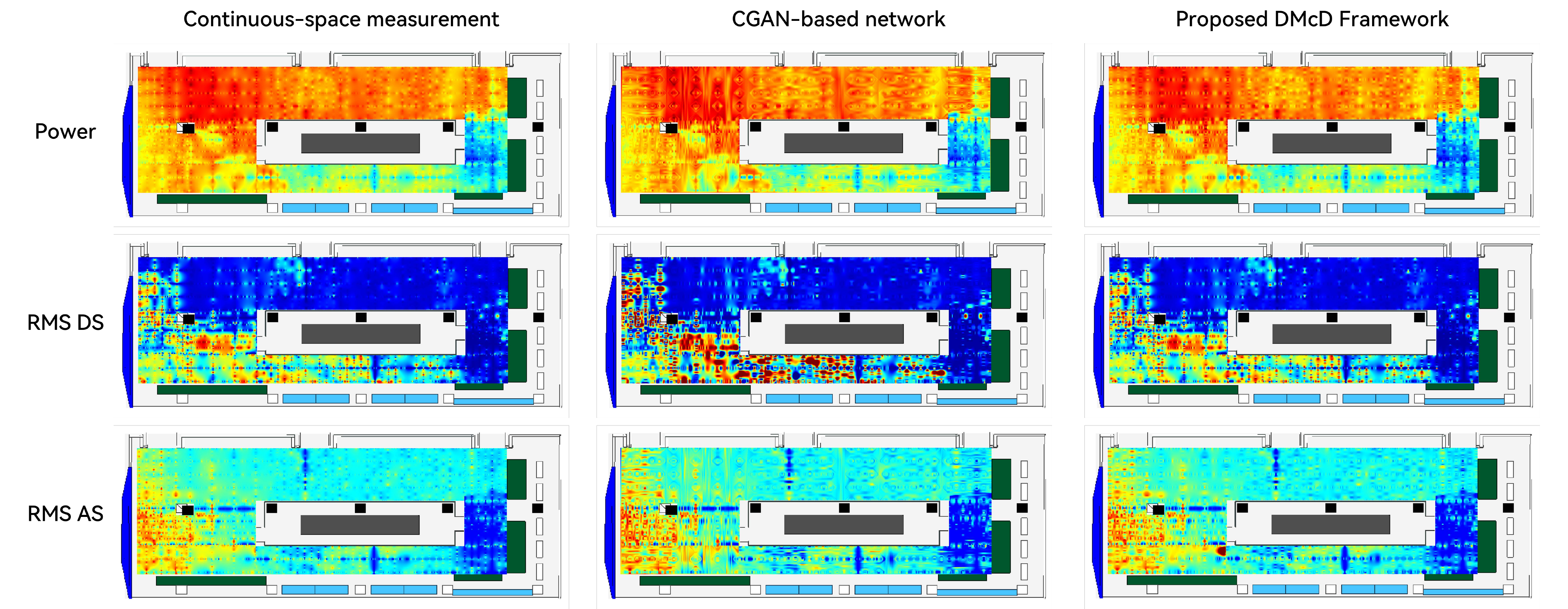}
	\caption{Channel maps for received power, RMS DS, and RMS AS: measurements baseline, CGAN-based method, and the proposed DMcD.}
		\vspace{-1em}
	\label{fig:Map_Comparison}
\end{figure*}

\subsection{Performance Evaluation of the Data-Model Co-Driven Channel Map}

\begin{figure}[t]
	\centering
	\includegraphics[width=9cm]{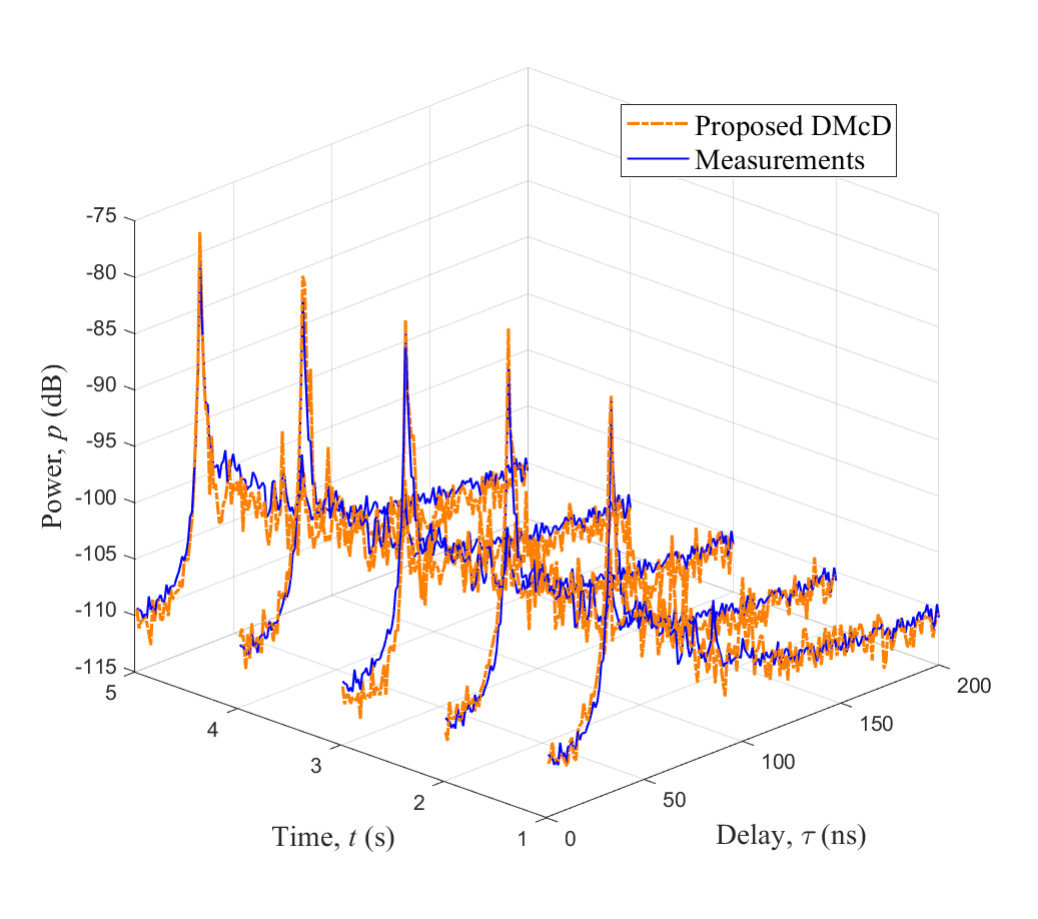}
	\caption{Time-varying delay PSD comparison between measurements and the proposed DMcD.}
	\vspace{-1em}
	\label{fig:Dpsd_True_vs_Predict}
\end{figure}

This subsection evaluates the end-to-end performance of the proposed DMcD framework.
The DMcD integrates the RT-based HCM for fast data augmentation and the proposed InductE-GNN for inductive interpolation, aiming to provide space--time-continuous channel maps with improved accuracy.

\subsubsection{Accuracy under indoor continuous-space measurements}

The indoor measurement scenario is static, and thus the evaluation focuses on the spatial accuracy at a fixed time slice. 
Fig.~\ref{fig:Map_Comparison} compares the channel maps of received power, RMS delay spread, and RMS angular spread obtained from the continuous-space measurements, a CGAN-based data-driven method, and the proposed DMcD framework. As shown in Fig.~\ref{fig:Map_Comparison}, the proposed DMcD produces power, DS, and AS maps that are visually consistent with the measurement-based reference over the whole area. 
In particular, in the NLoS regions behind the inner room and around the pillar, the CGAN-based results exhibit patchy artifacts and abrupt transitions, which are noticeable in both the power and DS maps. 
Such discontinuities are mitigated by the proposed DMcD, where smooth yet location-dependent variations are preserved, especially near the boundaries between LoS and NLoS zones.
A similar trend is observed in the AS maps. The CGAN-based method tends to introduce locally inconsistent angular dispersion patterns, while the proposed DMcD better follows the gradual spatial evolution shown in the measurements.

This improvement can be attributed to the joint design of DMcD. 
The HCM provides a physically consistent prior that already reflects MPCs birth-death and the diffuse scattering effects captured by the DMC component. 
Based on the prior, InductE-GNN performs residual refinement via topology-aware message passing, which suppresses unstable textures that may arise in purely data-driven method under discrete data.

\subsubsection{Accuracy under outdoor dynamic measurements}
Fig.~\ref{fig:Dpsd_True_vs_Predict} further evaluates the spatio-temporal consistency of the DMcD by comparing the time-varying delay PSD between the DMcD prediction and the reference data.
The predicted delay PSD follows the dominant components and the diffuse tail over time, and the overall pattern remains stable without noticeable drift.
This result indicates that once the time evolution is captured by the model-driven stage, the proposed inductive interpolation can reliably maintain the accuracy of channel maps during~online~updates.

\subsubsection{Update latency under online construction}
In addition to accuracy, the refresh latency of the proposed DMcD framework is evaluated for online channel map updates. The refresh period is 0.6~s, consisting of 0.4~s for HCM-based channel generation (model-driven stage) and 0.2~s for InductE-GNN inference (data-driven stage). The runtime is measured on a system equipped with an Intel Core i9-14900 CPU and 32~GB RAM. Admittedly, this latency may not be advantageous compared with AI-enabled map construction methods that target only a single channel attribute \cite{Wang2025SR,channelGAN,radiodiff}, due to their much lower output dimensionality. Nevertheless, DMcD provides a complete channel representation, jointly covering large-scale and small-scale channel information, thereby offering richer and more actionable channel awareness for embodied intelligent agents. With an update rate of approximately 1.7~Hz, the DMcD framework is already sufficient to support timely channel queries for embodied intelligence agents.
	
	\vspace{-0.5em}
\section{Conclusions}

	This paper has proposed a DMcD framework to construct a space-time continuous channel map, which can serve as a queryable wireless world model for embodied intelligent agents to achieve wireless channel perception. 
	The DMcD framework propose a two-stage channel map interpolation. 
	In the model-driven stage, a H-RT/GBSM has been developed to jointly capture static paths, dynamic scatterers, and DMCs, so that accurate time-varying CSI can be generated with measurement-consistent statistics. This stage acts to capture the impact of physical environmental changes on wireless channels while augmenting the dataset. In the data-driven stage, an InductE-GNN has been designed to further interpolate the channel map from dynamic priors and measurements. 
	Through aggregation functions and message passing mechanism, the InductE-GNN overcomes the prediction performance degradation caused by node feature shift, further achieving online interpolation.
	Evaluations in an indoor continuous-space scenario and an outdoor dynamic scenario have validated the key components of DMcD. The HCM matched the measured dispersion characteristics. The inductE-GNN achieved lower NMSE than baselines and showed stable interpolation performance under different sampling densities. Finally, the DMcD produced maps that better preserved continuity and location-specific channel structures, and tracked time-varying delay PSD of measurements well. 

 \bibliographystyle{IEEEtran}  
\bibliography{ref}

\end{document}